\documentclass[useAMS,usenatbib]{mn2e}
\usepackage{natbib}
\usepackage{color,graphicx} 
\usepackage{amsmath}        
\usepackage{amsfonts}       
\usepackage{amssymb}        
\usepackage{hyperref} 
\usepackage{subfig}
\usepackage{caption}
\usepackage{mathtools}

\usepackage{wasysym}

\title[Scintillation Noise in Exoplanet Photometry]{Atmospheric Scintillation Noise in Ground-Based Exoplanet Photometry}

\author[D F\"{o}hring et al.]{D. F\"{o}hring$^{1,2} $\thanks{E-mail: fohring@hawaii.edu},
R. W. Wilson$^{1}$, J. Osborn$^{1}$, V. S. Dhillon$^{3,4}$ \\
$^{1}$Department of Physics, Centre for Advanced Instrumentation, University of Durham, South Road, Durham, DH1 3LE, UK\\
$^{2}$Institute for Astronomy, University of Hawaii, 2680 Woodlawn Drive, Honolulu, HI 96822, USA\\
$^{3}$Department of Physics and Astronomy, University of Sheffield, Sheffield S3 7RH, UK\\
$^{4}$Instituto de Astrof{\'i}sica de Canarias, 38205 La Laguna, Tenerife, Spain\\}

\begin{document}
\maketitle

\begin{abstract}
Atmospheric scintillation caused by optical turbulence in the Earth's atmosphere can be the dominant source of noise in ground-based photometric observations of bright targets, which is a particular concern for ground-based exoplanet transit photometry. We demonstrate the implications of atmospheric scintillation for exoplanet transit photometry through contemporaneous turbulence profiling and transit observations. We find a strong correlation between measured intensity variations and scintillation determined through optical turbulence profiling. This correlation indicates that turbulence profiling can be used to accurately model the amount of scintillation noise present in photometric observations on another telescope at the same site. We examine the conditions under which scintillation correction would be beneficial for transit photometry through turbulence profiling, and find that for the atmosphere of La Palma, scintillation dominates for bright targets of magnitude above $V\sim10.1$~mag for a 0.5~m telescope, and at $V\sim11.7$~mag for a 4.2~m telescope under median atmospheric conditions. Through Markov-chain Monte Carlo methods we examine the effect of scintillation noise on the uncertainty of the measured exoplanet parameters, and determine the regimes where scintillation correction is especially beneficial. The ability to model the amount of noise in observations due to scintillation, given an understanding of the atmosphere, is a crucial test for our understanding of scintillation and the overall noise budget of our observations.
\end{abstract}


\section{Introduction} \label{intro}

Atmospheric scintillation significantly degrades the quality of time-resolved photometric data that can be obtained from ground-based observations compared to those from space.
With the number of ground-based transit surveys increasing (e.g. \citealt{snellen2012, wheatley2013}), as well as the prevalence and necessity for ground-based follow-up of exoplanet transits for characterisation, it is important to investigate the effects of scintillation on exoplanet light curves.

The decrease in flux during an exoplanet transit is small compared to the flux of the host star: around 1 per cent for the transit of a Sun-Jupiter system and 0.1 per cent for its secondary eclipse. Bright stars are ideal exoplanet host star candidates; they can be found more easily by wide field surveys and provide adequate flux for high-resolution spectroscopic follow-up. For this reason, several of the next-generation of transit searches will focus on finding planets around stars of magnitude 13 or brighter, such as the Next-Generation Transit Survey \citep{wheatley2013} or the Multi-site All-Sky CAmeRA \citep{snellen2012}. However, scintillation sets a fundamental limitation on the precision that can be obtained for exoplanet transit photometry from the ground. While adaptive optics has been successfully used to provide diffraction-limited imaging, the first generation of instruments for correcting for the intensity fluctuations caused by scintillation are only currently in development (\citealt{osborn2011, viotto2012,  osborn2015, dhillon2016}). For these instruments, understanding the regimes where scintillation is the dominant source of noise is essential. 

Optical scintillation has been previously described by \citet{roddier1981}, \citet{dravins1997a} and \citet{osborn2015}. Previous work on the detection limits of fast photometry has been performed by \citet{mary2006} and \citet{southworth2009}, who break down noise on a light curve into constituent parts and investigate the relative contributions of scintillation and photon noise through statistical analysis of scintillation.  However, these calculations are based on Young's equation of scintillation noise \citep{young1967}, which assumes an atmosphere with an averaged turbulence profile that may not be correct for any one particular given time. Measurements of atmospheric turbulence using, for example, Stereo-SCIDAR (SCIntillation Detection And Ranging) \citep{shepherd2014} have shown that a more accurate representation of atmospheric turbulence is one which consists of a number of discrete layers displaying Kolmogorov statistics. Young's formula underestimates scintillation noise by a factor of 1.4 -- 1.6 on average, depending on the site \citep{kornilov2012}. Likewise, scintillation estimates using Stereo-SCIDAR on La Palma show a much greater range of scintillation noise with higher average than would be expected from Young's approximation \citep{osborn2015}. In addition to the differences between sites,  atmospheric turbulence and wind velocity profiles obtained using Stereo-SCIDAR show that optical turbulence evolves rapidly, on a minute to minute time scale, so the actual scintillation noise can vary considerably from night to night, depending on the magnitude of the high altitude turbulence. 

Turbulence profiling is used to determine the refractive index structure function of the atmosphere, $C_n^2(h)$, a measure of the optical turbulence strength. Observations made using Stereo-SCIDAR \citep{osborn2015obs} additionally give the velocity of the individual turbulent layers which is needed to estimate scintillation. The stereo-SCIDAR is an optical triangulation technique, described in detail in \citet{osborn2013} and \citet{shepherd2014}, based on SCIDAR (SCIntillation Detection And Ranging) \citep{klueckers1998}, \citep{fuchs1998} and the instrument design of the Conjugate-Plane Photometer by \citet{osborn2011}. It measures turbulence strength and wind direction by imaging optical binary stars, with an angular separation of $\theta$, on two separate detectors, and calculating the covariance of the intensity patterns for each star. The amplitude of the peaks in of the spatial cross covariance functions of the images is related to the turbulence strength, and their separation correspond to $h\theta$, where $h$ is the altitude of the turbulent layer. Wind velocities are obtained by measuring the vectors between the temporal cross-covariance peaks of consecutive pupil images. 

For exposure times $t_\textrm{exp}$ longer than the wind crossing time given by $D / V_\perp$, where $D$ is the diameter of the telescope primary mirror, an estimate of intensity variations due to scintillation for discrete layers, $\sigma_I^2$, requires knowledge of $V_\perp(h)$, the vertical profile of the perpendicular wind velocity. For discrete layers, it is given by \citet{kenyon2006} as
\begin{equation}
\sigma_I^2 = 10.7 \, \textrm{cos}(Z)^{-3.5}\int^\infty_0 \frac{C^2_n(h) h^2}{V_\perp(h)} \textrm{d}h D^{-4/3} t_\textrm{exp}^{-1},
\label{kenyon}
\end{equation}
where $Z$, is the zenith angle, $h$ the height of the turbulent layer.
Scintillation arises from high altitude atmospheric turbulence, which has a horizontal extent typically of a few tens of km (e.g. \citet{vinnichenko2013}). For this reason, to first order, noise level is expected to be isotropic within the same site, with only small differences arising due to differing lines of sight. This allows for concurrent monitoring of scintillation noise using an optical turbulence profiler, such as Stereo-SCIDAR \citep{shepherd2014}. Such profiling will be useful for estimating the error-budget of the next generation ground-based transit surveys. 

With new scintillation correction techniques, such as conjugate-plane photometry and tomographic atmospheric reconstruction \citep{osborn2011, osborn2015}, a reduction of scintillation noise to the level of the photon noise will be possible, depending on the state of the atmosphere. This will result in an improvement in the uncertainties of the measured exoplanet transit parameters. 

In this paper, we present the results of concurrent turbulence profiling using Stereo-SCIDAR and exoplanet transit observations carried out using the $pt5m$ telescope  \citep{hardy2015} and ULTRACAM \citep{dhillon2007} on the William Herschel Telescope (WHT) on La Palma in section \ref{obs}. The aim is to show that turbulence profiling may be used for scintillation characterisation for exoplanet photometry. The regimes where scintillation is a limiting source of noise on photometry are examined in the $V$-band in section \ref{scintlim}, and its effect at near infrared wavelengths is discussed in section 
\ref{wave}. The extent to which scintillation noise on the light curve affects the uncertainty of the measured astrophysical parameters for the transit and secondary eclipse is examined by fitting synthetic light curves with and without scintillation noise through Markov-Chain Monte Carlo (MCMC) methods in section \ref{modelling}. Finally our findings are summarised and discussed in section \ref{disc}.
\section{Observation and Analysis} \label{obs}

To investigate how well the error budget of photometric measurements are accounted for when all the scintillation noise is known, three nights of contemporaneous turbulence profiling and photometry were carried out using Stereo-SCIDAR on 2013 July 20 on the 1.0m Jacobus Kapteyn Telescope and between 2014 March 15 -- 16 on the 2.5~m Isaac Newton Telescope (INT). Photometric observations were made using ULTRACAM  at the 4.2 m William Herschel Telescope (WHT) and the 0.5 m \textit{pt5m}. The scintillation noise modelled from the SCIDAR data was compared to the measured intensity variance of the transit light curves as described below.

\subsection{Scintillation Measurements from Stereo-SCIDAR} \label{obs}

Scintillation noise was calculated from the Stereo-SCIDAR turbulence profiles using equation \ref{kenyon}. Variables relating to the atmosphere were extracted from Stereo-SCIDAR and combined as $\sum C_n^2(h) h^2 / V_\perp$ for ease of comparison. This parameter was afterwards scaled according to the telescope diameter, object brightness and exposure time for each science observation. Due to the method used, wind velocity measurements were only possible when the peaks in temporally adjacent pupil images could both be distinctively resolved. This means that only the strongest layers, accounting for approximately six per cent of the total $C_n^2$ measurements had velocities associated with them. The rest of the $C_n^2$ measurements were assigned an average velocity within a given layer. These layers were determined by combining all the wind velocity estimates for an entire night and sorting them by height. Layer boundaries were  placed if no wind velocity is detected for 100 m above the last measured velocity. Typically, around 40 layers were distinguished each night. During the course of a night the wind velocity for a given height was assumed constant. Anomalous wind velocities resulting from difficulty of the Stereo-SCIDAR algorithm to detect cross-covariance peaks were excluded during a second pass and replaced by the average of the neighbouring four points. These approximations were deemed valid, as scintillation mainly arises from the strongest layers which were likely to have corresponding measured velocities.

\begin{figure*}
  \includegraphics[width=7.5in]{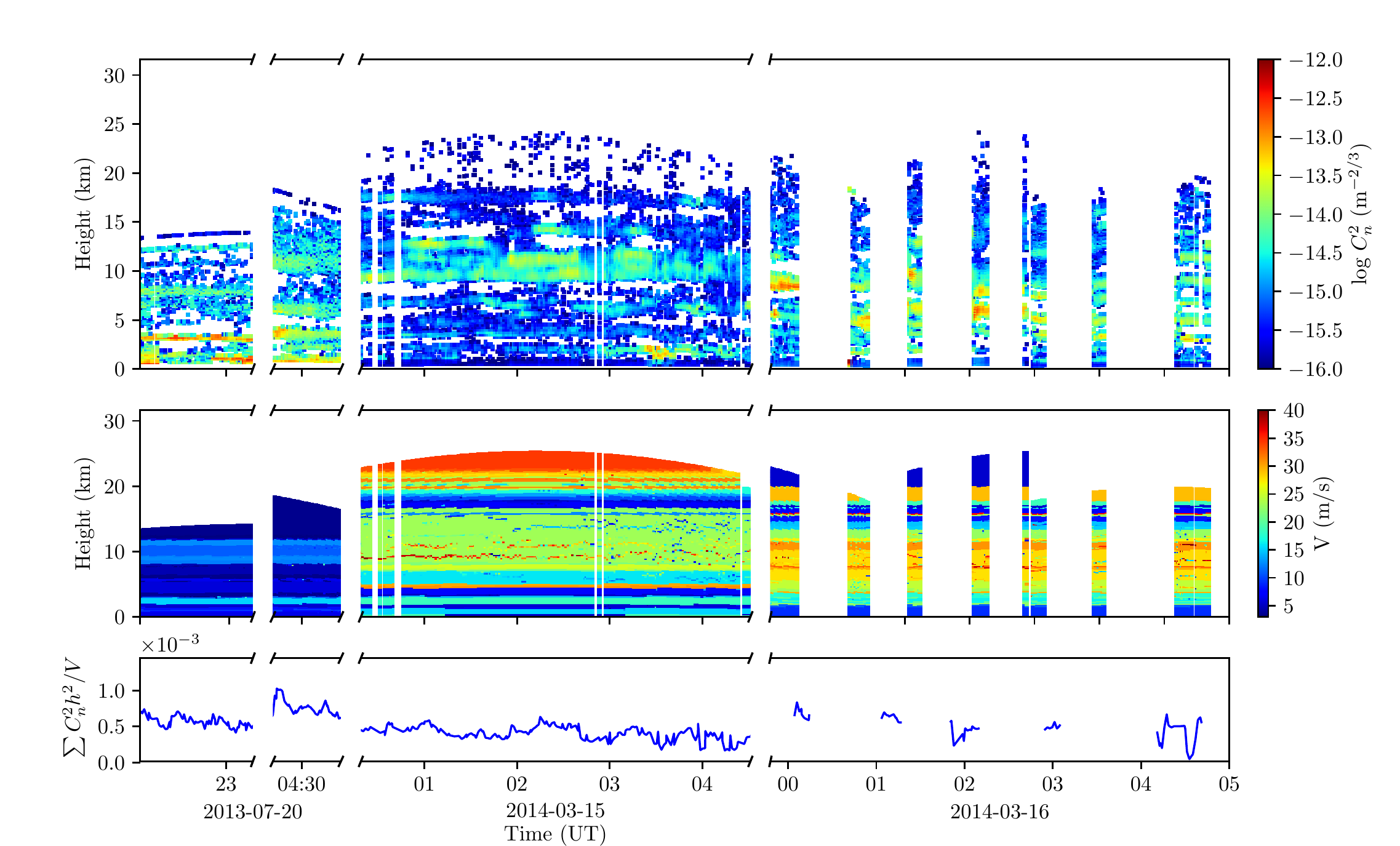}
  \caption{Atmospheric data obtained using Stereo-SCIDAR. The top panel shows the measured $C_n^2$ profile with height. Data points with $C_n^2$ weaker than $10^{-16}$m$^{-2/3}$ have been removed for clarity. The middle panel shows the determined average wind velocity for the layers. The bottom panel shows the sum of $C_n^2(h) h^2 / V_\perp$ calculated from the information in the top two panels.}
  \label{sciprof}
\end{figure*}

 The profiles and values for $\sum C_n^2(h) h^2 / V_\perp$ obtained from stereo-SCIDAR are shown in figure \ref{sciprof}.
 
\subsection{Noise Measurements on ULTRACAM Photometry}

The photometry from ULTRACAM and the \textit{pt5m} telescope were corrected for bias and flat-field, and aperture photometry was performed using the ULTRACAM pipeline written by Tom Marsh\footnote{\url{http://deneb.astro.warwick.ac.uk/phsaap/software/ultracam/html/index.html}}. The differential flux of the brightest stars in the field for each data set was obtained in the $r'$-band by optimising the aperture radii to produce the light curve with the least amount of scatter in the out-of-transit region. If there was more than one bright potential comparison star in the observed field, the stars that produced the most stable differential flux (lowest ratio of $\sigma/\mu$) were chosen and the uncertainties from photon counting and system noise were combined in quadrature. Not all data obtained contained a transit, some were bright binaries or exoplanets not undergoing a transit at the time. If the data contained a transit, a full light curve fit was performed and the data was divided by the fit. This was necessary, as otherwise the presences of the transit would have produced a systematic offset in the standard deviations. The light curve was calculated using the \citet{mandel&agol} analytical equations with quadratic limb darkening with 6 free parameters for the planetary and stellar radii, inclination, limb darkening coefficient, offset from predicted transit mid-time and two airmass coefficients and fit using a downhill simplex method. The total photometric noise, $\sigma_m$ was calculated by measuring the normalised standard deviation of the differential counts in a running interval of 300~s. 

The measured total photometric noise was compared to the total modelled noise estimate, which consists of the scintillation noise modelled from the SCIDAR data and the known internal noise of the ULTRACAM system. The ULTRACAM noise estimates are obtained as an output from the reduction pipeline and include photon noise from the sky and object, and a small amount of detector noise.

\subsubsection{2013 July 20}

\begin{figure*}
   \centering
   \includegraphics[width=7.5in]{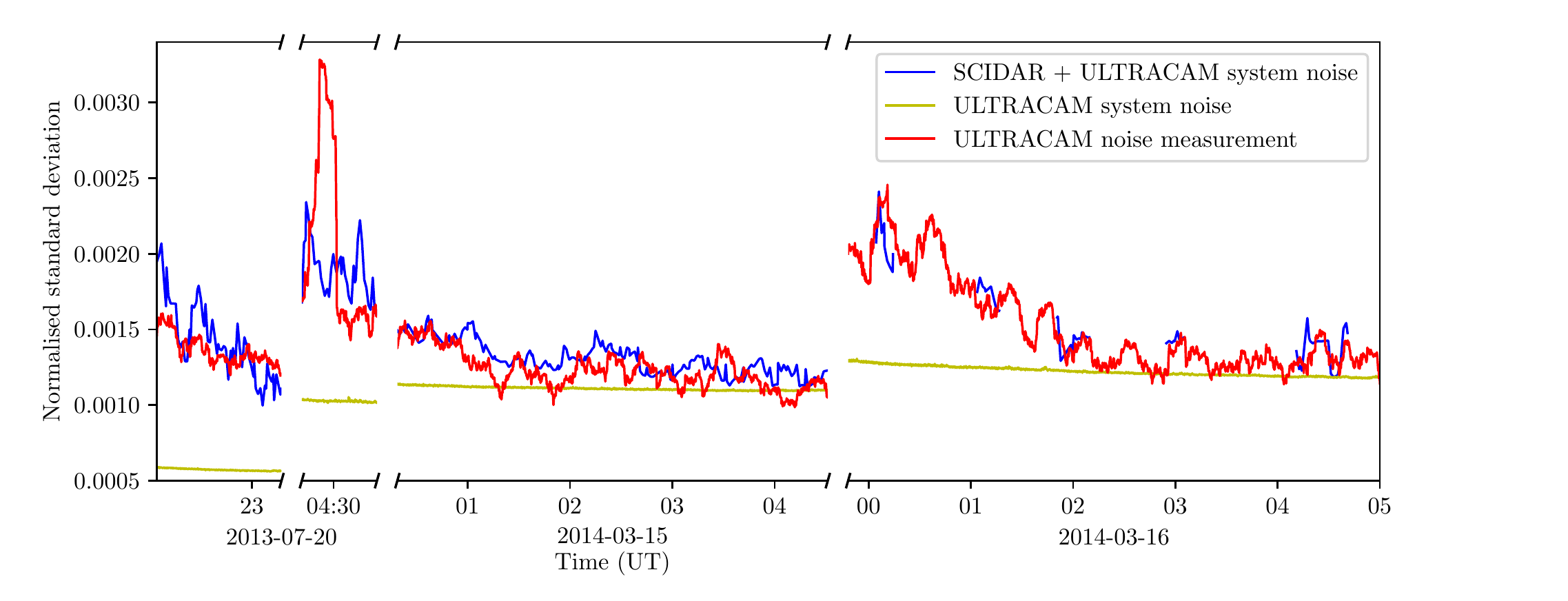}
   \caption{Comparison between modelled noise from SCIDAR and measured noise from ULTRACAM. The red is line shows the normalised standard deviation of the measured flux from ULTRACAM. The modelled noise (blue) takes into account he scintillation noise from the SCIDAR data, which is scaled according exposure time, telescope diameter and airmass of the observation, and the internal system noise from ULTRACAM. The internal ULTRACAM noise includes photon noise and the small amount of detector noise, shown in yellow. The different levels of the yellow line is due to different targets being observed at different times, which contribute different amounts to the photon noise.}
   \label{1}
\end{figure*}

The ULTRACAM photometry on the night of 2013 July 20 consisted of two fields at the beginning and end of the night. The first field contained HAT-P-23 at RA 20:24:29.72 Dec +16:45:43.8, and two comparisons TYC 1632-1319-1 and TYC 1632-1019-1, with $V$-magnitudes of 11.4 and 12.3 respectively. The second contained two fainter, $V \sim 13$ stars in the field around the target 2MASS 23043114+4927344 at coordinates RA 23:04:45.76 Dec +49:26:24.8 and RA 23:05:18.67 Dec +49:27:10.3. Both sequences were taken with exposure times of 0.5~s. Figure \ref{1} shows the comparison between the measured and modelled noise for the $r'$-band photometry. On this night the length of contemporaneous ULTRACAM photometry is short, but the data appear overall in close agreement. Discrepancies between Stereo-SCIDAR and ULTRACAM noise are caused by both telescopes pointing in different directions.  

\subsubsection{2014 March 15}

\begin{figure}
   \centering
   \includegraphics[width=3.5in]{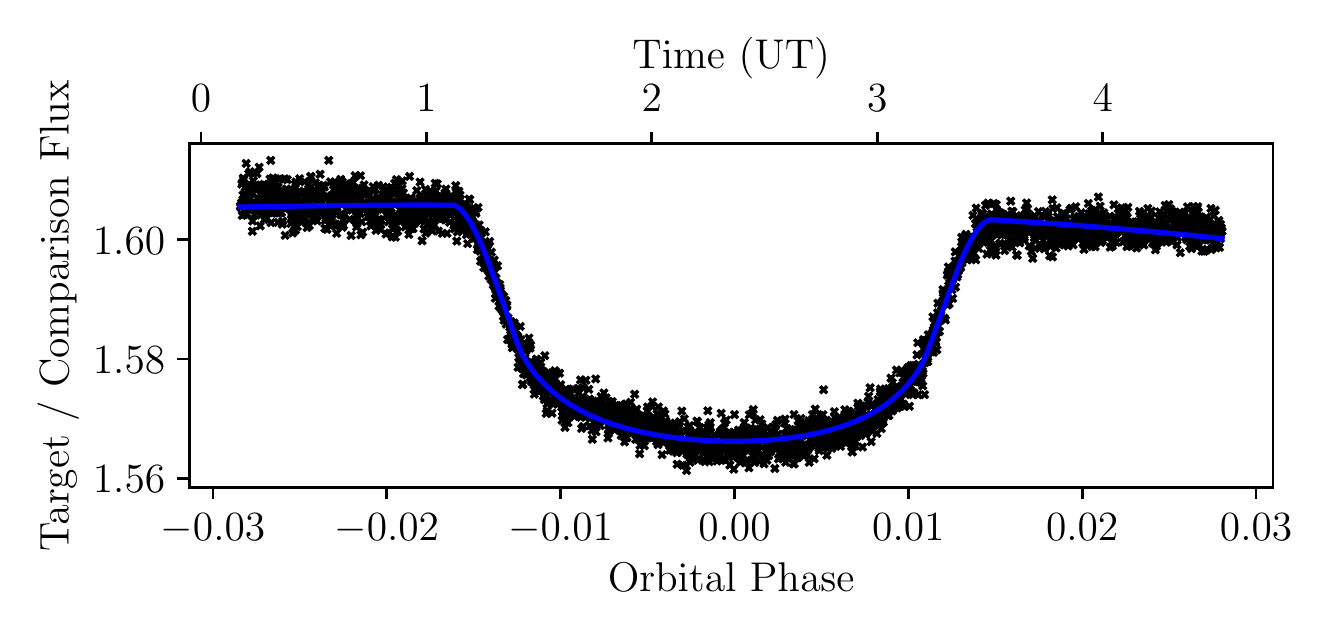}
   \caption[Light curve and fit of the transit of HAT-P-12b in the $r'$-band on the night of 2014 March 15 using ULTRACAM on the 4.2m WHT]{Light curve and fit of the transit of HAT-P-12b in the $r'$-band on the night of 2014 March 15 using ULTRACAM on the 4.2 m WHT.}
   \label{2lc}
   \end{figure}

On the night of 2014 March 15 the transit of HAT-P-12b was observed at the WHT. The target and comparison were HAT-P-12, a $V = 12.8$~mag star and a nearby star at RA 13:57:24.96 Dec +43:31:33.4 of similar brightness, respectively. Throughout the night the seeing was stable, around 1-1.5". Stereo-SCIDAR measurements show that the turbulence was close to the ground, resulting in very little scintillation. The exposure time was 0.5~s. Figure \ref{2lc} shows the light curve fitted to the data obtained for the transit in the $r'$-band.

The measured scintillation is the same order of magnitude as the scintillation noise modelled from the SCIDAR data and appears correlated, as shown in figure \ref{1}. The measured standard deviation of the differential flux (red line) on occasion dips below the expectation for the system noise (yellow line) due to differences in the line of sight between the two telescopes, but on average lies above or at the system noise level, which does not include scintillation. The fractional contribution from scintillation is low, due to the combination of weak turbulence and relative faintness of the targets. The data is photon noise dominated, with scintillation noise around 20$\%$. 

\subsubsection{2014 March 16}

\begin{figure}
   \centering
   \includegraphics[width=3.5in]{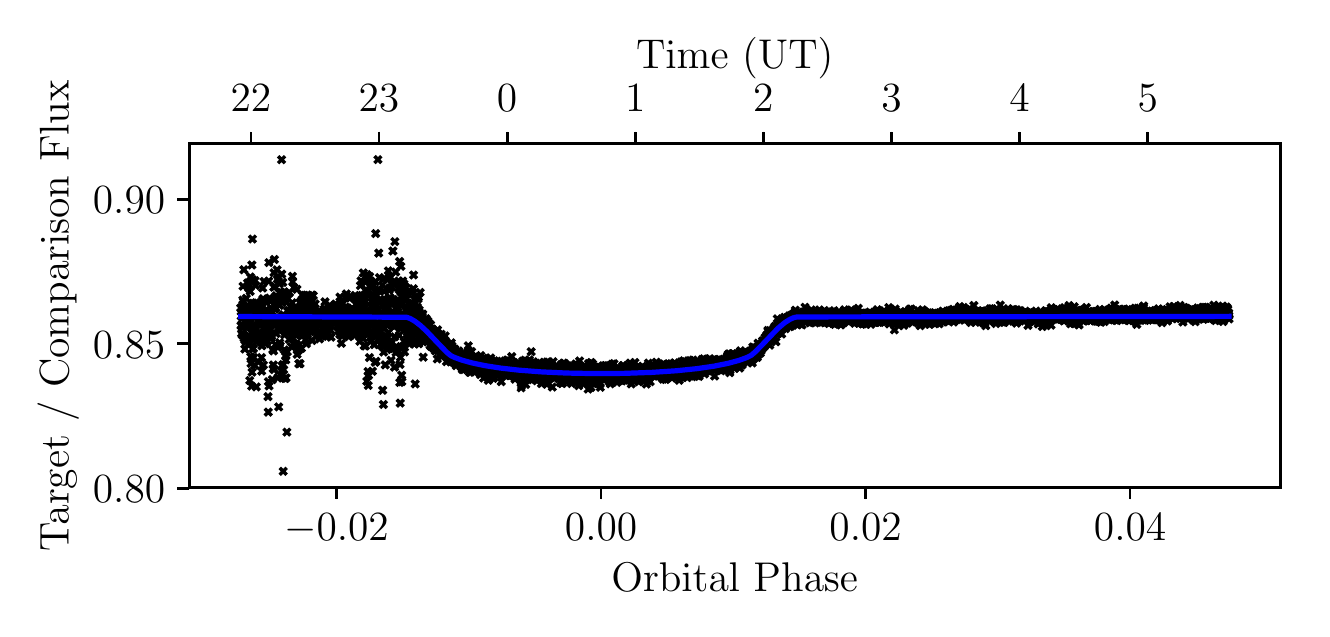}
   \caption{Light curve and fit of the transit of HAT-P-44b in the $r'$-band on the night of 2014 March 16 using ULTRACAM on the 4.2m WHT.}
   \label{3lc}
\end{figure}

\begin{figure}
   \centering
   \includegraphics[width=3.5in]{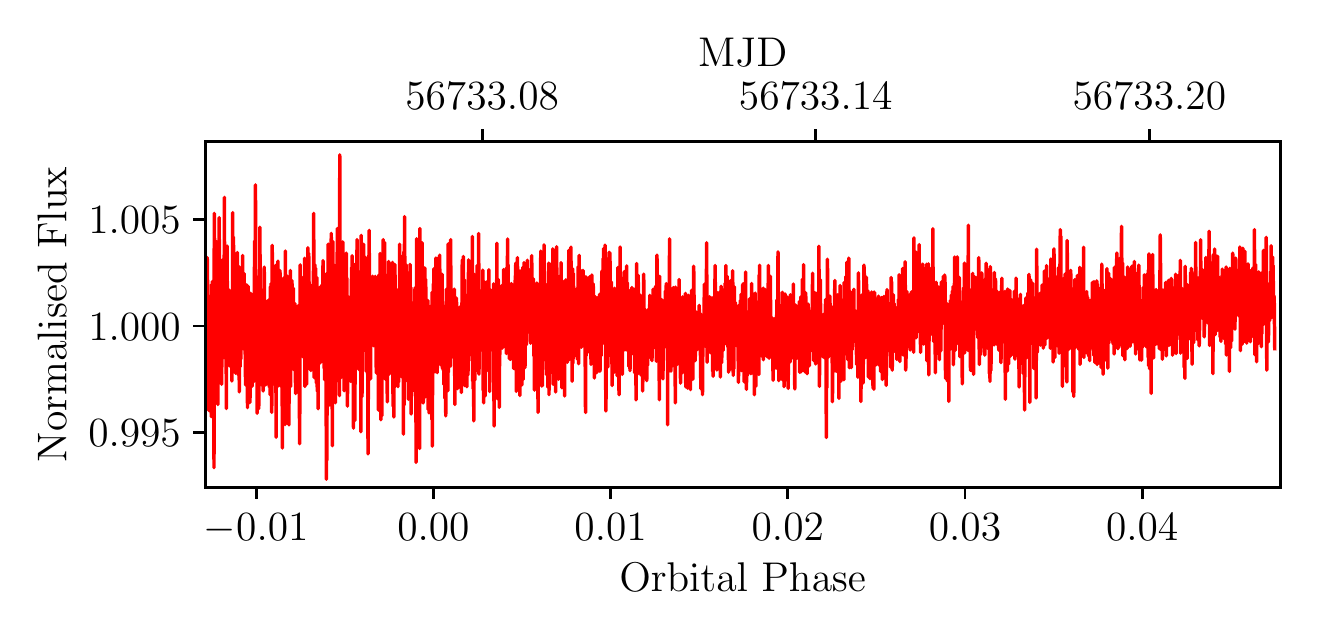}
   \caption[Differential photometry of HAT-P-44b with the transit removed]{Differential photometry of HAT-P-44b with the transit fitted and divided out. The portion of the light curve with transparency variations is not included in this plot, to emphasise the improvement in the scatter of the data with time caused by scintillation.}
   \label{3corr}
\end{figure}

On the night of 2014 March 16 the transit of HAT-P-44b was observed using ULTRACAM while the out-of-transit light curve of WASP-54 was observed using the $pt5m$ telescope. The light curve from ULTRACAM was constructed from the $V = 13.2$~mag star GSC 03465-00123 (HAT-P-44) and a nearby star of similar brightness at RA 14:12:42.10 Dec +47:01:05.0. The light curve from the $pt5m$ was created from the bright, $V = 11.0$~mag target, BD+00 3088 (WASP-54) and a nearby comparison HD 119217. Figure \ref{3lc} shows the light curve and fit obtained for HAT-P-44. The data with the light curve removed (figure \ref{3corr}) shows an improvement in scatter as the night progresses, which is reflected in figure \ref{1}. 

\begin{figure}
   \centering
   \includegraphics[width=3.5in]{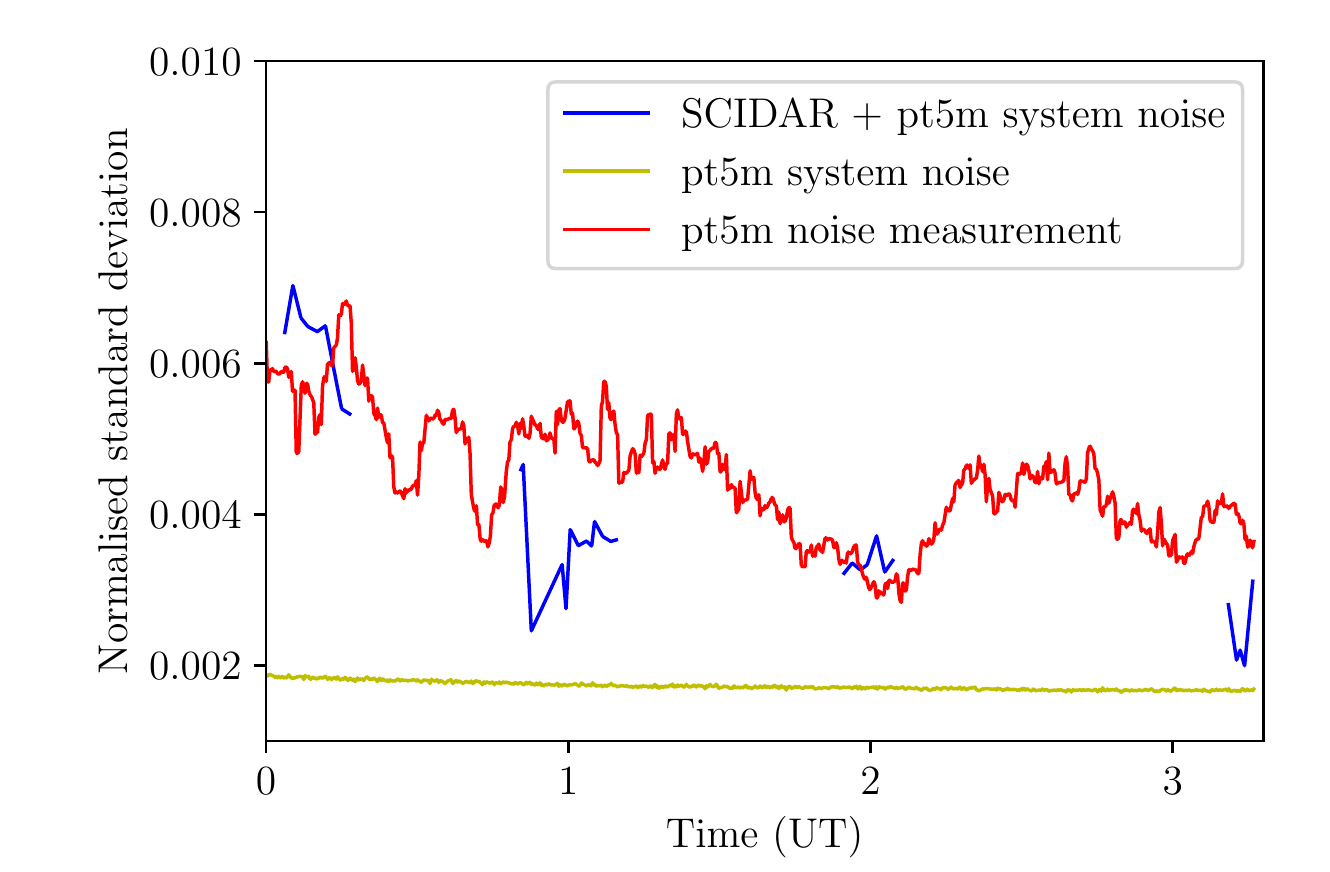}
   \caption[Comparison between predicted and measured scintillation from $pt5m$ on 2014 March 16]{As figure \ref{1} for the $pt5m$ on 2014 March 16.}
   \label{3c}
\end{figure}

The comparison between the observed and modelled standard deviation for the $pt5m$ observations is shown in figure \ref{3c}. The modelled noise matches the measured noise on the light curve well and the same improvement with time is observed on both light curves.

\subsection{Comparison between Stereo-Scidar noise estimate and ULTRACAM noise measurements}

\begin{figure}
   \centering
   \includegraphics[width=3.5in]{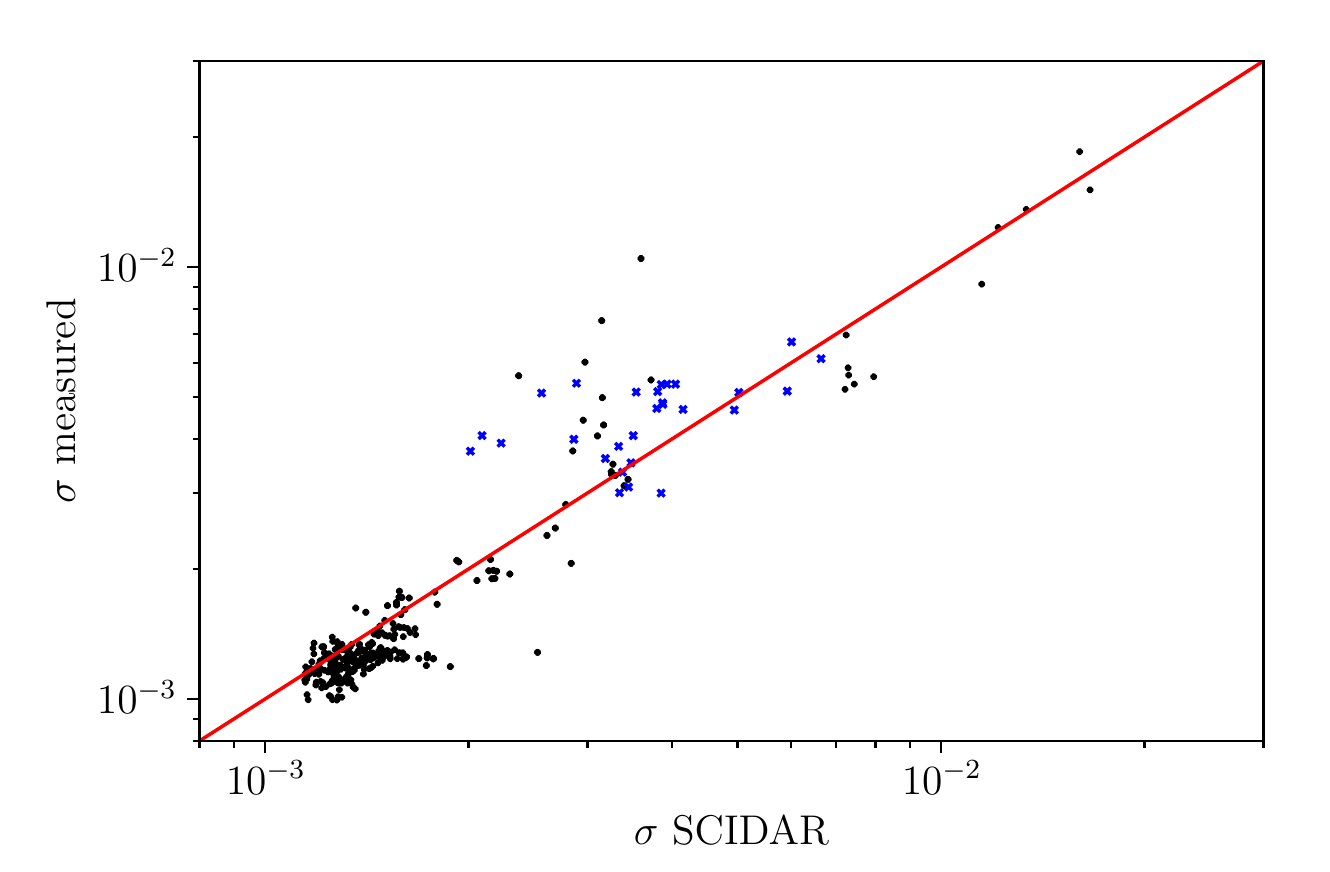}
   \caption{Comparison of measured noise on observation and estimated noise from Stereo-SCIDAR profiles. The dots represent data points from ULTRACAM and the crosses represent data points from the $pt5m$.}
   \label{scidarlog}
\end{figure}

Combining a total 348 independent observations made with two telescopes over three different nights, we found that the noise level modelled from Stereo-SCIDAR observations and measured noise agree with a correlation of 0.93 (figure \ref{scidarlog}). Compared to estimates of scintillation noise from Young's equation, where the agreement is only 0.63. The strong correlation means that through knowledge of the turbulence profile at any given instant we were able account for the majority noise budget of these photometric observations, even when the contribution of scintillation was small. This indicates that most of the noise other than photon and instrumental noise on the observations was due to scintillation, and when scintillation noise was correctly accounted for, the amount of noise that was needed to be attributed to additional systematic or `red' noise was much less. Differences between the Stereo-SCIDAR turbulence profile measurements and observed scintillation may arise due to the differing line of sight of the telescopes, so that localised changes in the instantaneous turbulence strength and wind velocity can cause the scintillation estimates to be out of agreement.  
 
\section{Is scintillation correction worthwhile for exoplanet transit observations?} \label{scintlim}

The conditions under which scintillation correction is worthwhile for high-precision time-resolved photometry are examined in this section. Comparing the noise due to scintillation to the shot-, or photon noise, $\sigma_s$, one can calculate the limiting magnitude at which scintillation becomes the dominant source of noise on the observation. For stars fainter than the limiting magnitude, scintillation will be less significant, compared to the other sources of noise.

Atmospheric scintillation correction is the most useful for bright stars, where it is the dominant source of noise on an observation. For modern instruments observing bright transits, the noise due to sky background and detector readout are typically negligible compared to the shot noise from photon counting of the source flux. The fractional shot noise, $\sigma_s$,  is given by $\sqrt{N} / N $, where $N$ is the number of photons. For a star of apparent magnitude $m$, this may be calculated as 

\begin{equation}
\sigma_s^2 = \left(  \phi ~ \Delta \lambda ~A ~ E ~ t_\textrm{exp}   / ~2.5^m \right)^{-1},
\label{photnoise}
\end{equation} 
where $\phi$ is the photon flux in photons s$^{-1}$ cm$^{-2}$ \AA$^{-1}$, $\Delta \lambda$ is the spectral response range of the detector, $A$ is the light collecting area of the telescope given by $\pi ~(D/2)^2$ and $E$ is the efficiency, a combination of the CCD quantum efficiency and the combined throughput of the atmosphere and instrument.

While scintillation is the result of the sum of turbulent layers of varying $C_n^2(h)$, for ease of comparison, and to produce a statistical description of scintillation we define a parameter, $\sum C_{n, 10\textrm{km}}^{'2}$, the equivalent turbulence strength of a single layer at a height of 10~km that gives rise to the same measured scintillation index as the sum of the scintillation indices resulting from every individual layer at each different altitude. The value of 10~km is chosen as the boundary of the Troposphere, the origin of most of the atmospheric scintillation. The limiting magnitude can thus be calculated for a theoretical telescope by setting the noise due to scintillation described by equation \ref{kenyon} equal to the fractional shot noise in equation \ref{photnoise} and rearranging:

\begin{equation}
m = \log\left(10.7 \frac{ \sum C_{n, 10\textrm{km}}^{'2}  h^2} {\overline{V}_\perp}\pi(50)^2 D^{2/3} \phi \Delta \lambda ~ E \right) \\ / \log(2.5).  
\label{limitingmag}
\end{equation} 

\begin{figure}
   \centering
   \captionsetup[subfigure]{oneside,margin={0.6cm,0cm}}
   \subfloat[]{\includegraphics[width=2.2in]{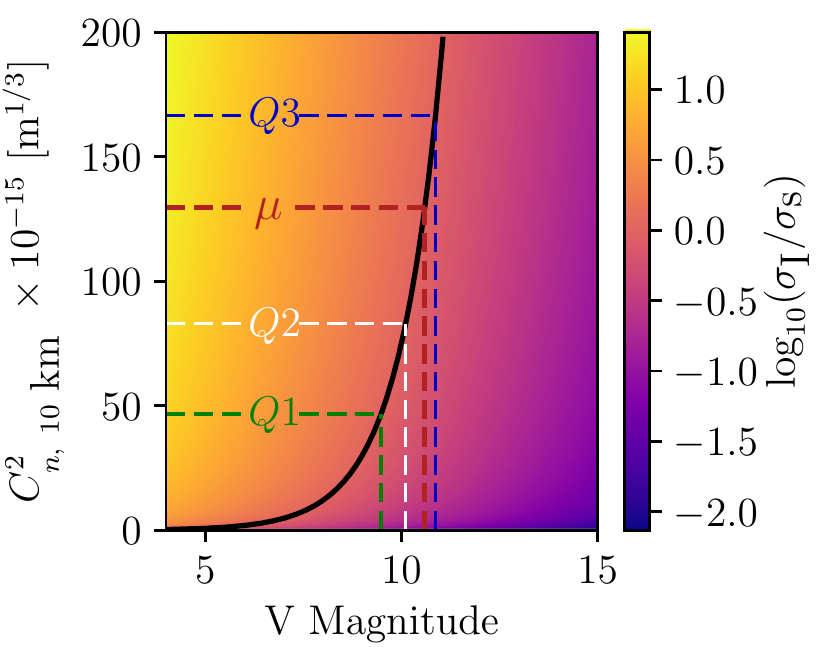}}\\
   \subfloat[]{\includegraphics[width=2.2in]{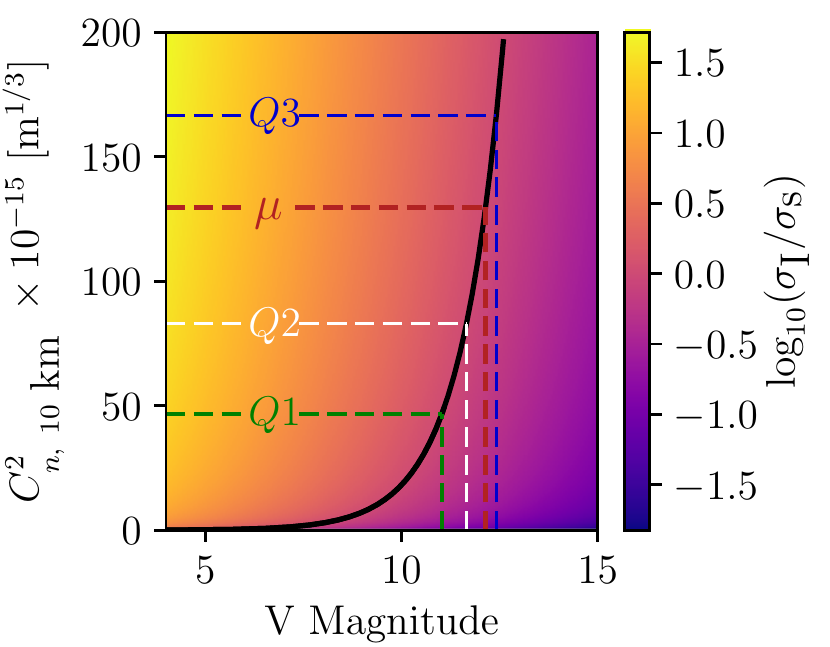}}
   \caption[Ratio of scintillation noise to photon noise for 0.5 m and 4.2 m telescopes]{Ratio of scintillation noise to photon noise for a 0.5 m (a) and 4.2 m (b) telescope, respectively, for varying $V$ magnitude and $\sum C_{n, 10\textrm{km}}^{'2}$. The dashed lines show the limiting magnitude for the mean, first, second and third quartiles of turbulence strength. The limiting magnitudes for median (Q2) La Palma seeing are $V = 10.1$~mag for a 0.5~m and $V = 11.7$~mag for a 4.2~m diameter telescope.}
   \label{limiting}
\end{figure}

Table \ref{table:lapalma} shows the mean, $\mu$, median, Q2, and first and third quartiles, Q1 and Q3, of the turbulence distribution of $\sum C_{n, 10\textrm{km}}^{'2}$ measured from 20 nights of turbulence profiling using Stereo-SCIDAR for La Palma. Assuming typical values for this type of observation, $E=0.4$, $\Delta \lambda = 900$~\AA $ $ and $ \overline{V}_\perp = 15~\textrm{ms}^{-1}$, enables the limiting magnitudes and the percentage contribution of scintillation to the overall noise budget to be calculated under different conditions. Columns 2 and 3 in table \ref{table:lapalma} show a summary of the limiting magnitudes for a theoretical small, 0.5~m and medium-size, 4.2~m telescope respectively.  Figure \ref{limiting} illustrates the fractional scintillation noise for a 0.5~m and 4.2~m telescope respectively, for the $V$-band.

\begin{table} 
\vspace{0.5cm}
\centering 
\begin{tabular}{l c c c} 
\hline\hline 
Quartiles 					& $\sum C_{n, 10\textrm{km}}^{'2}$   & limiting		& limiting\\ 
						&  ($\times 10^{-15}$ m$^{1/3}$)	& magnitude    & magnitude   \\
						&         						& for 0.5~m 	& for 4.2~m					\\
						[0.5ex] 
\hline 
Q1 & 46.7  & 9.5 & 11.0  \\ 
Q2 & 83.0  & 10.1 & 11.7 \\ 
$\mu$ & 129.8 & 10.6 & 12.1 \\
Q3 & 166.7  & 10.9 & 12.4  \\  [1ex]
\hline 
\end{tabular} 
\caption[Distribution of $\sum C_{n, 10\textrm{km}}^2$ for La Palma]{Distribution of $\sum C_{n, 10\textrm{km}}^{'2}$ for La Palma.  The rightmost column shows the corresponding magnitude limits in the $V$-band below which scintillation dominates.}
\label{table:lapalma} 
\end{table}
 
There is no exposure time dependence, as scintillation and photon noise both scale by the power of $t_{\textrm{exp}}^{-1/2}$. The results show that on La Palma, for an observing set-up typical for exoplanet photometry, scintillation dominates for bright targets of magnitude above $V = 10.1$~mag on for a 0.5~m telescope, and at $V = 11.7$~mag for a 4.2 m telescope under median atmospheric conditions. For fainter stars correction is still worthwhile: under median conditions, at $V = 11.2$~mag on a 0.5~m telescope, and $V = 12.8$ mag on a 4.2~m telescope, scintillation accounts for one third of the noise on the observation. At $V = 12.6$~mag and $V = 14.2$~mag, for $D = 0.5$~m and $D = 4.2$~m respectively, scintillation only contributes 10\% of the noise on the observation. Therefore, scintillation correction is expected to bring a benefit for photometry of stars that are brighter than these limits, on average. However, we stress that atmospheric turbulence varies considerably from night to night, and instantaneous limiting magnitude will vary case-by-case.

The limiting magnitudes calculated in this manner are valid for observations for La Palma. Turbulence measurements for other sites based on \citet{kornilov2012}, show similar limits for Maunakea, both being the two sites which have amongst the lowest scintillation noise levels in the world. For other sites, the threshold where scintillation dominates is fainter, but of comparable order. For example, at Paranal, where the median measured $\sum C_{n, 10\textrm{km}}^{'2} = 158 \times 10^{-15}$ m$^{1/3}$, the limiting magnitude for scintillation is $V = $10.8~mag for 0.5~m and $V = $12.4~mag for 4.2~m telescopes, respectively. 

\subsection{Wavelength Dependence of Scintillation} \label{wave}
 
Since much of exoplanet follow-up for characterisation is carried out in the infrared, we examine the effect of scintillation on observations at infrared wavelengths. As in section \ref{scintlim}, we do this by calculating the limiting magnitudes for scintillation, but here we take into account the increasing sky brightness towards the infrared.
Scintillation is independent of wavelength when the telescope diameter is much greater than the Fresnel radius, $r_F$, defined as $r_F \approx (\lambda h)^{1/2}$, where $h$ is the height of the turbulent layer and $\lambda$ is the wavelength \citep{dravins1997b}. This value is typically of the order of a few cm for observations both in the visible and infrared. However, in the infrared, thermal emission from the telescope, instrument and sky contribute significantly to noise and cannot be ignored as for optical observations. In the $J$-band the sky is around 17 times brighter and in the $K$ and $H$-bands it is around 400 times brighter than in the optical, so one would expect the impact of scintillation to be much less. Previous measurements, by e.g.\ \citet{wainscoat&cowie1992} and \citet{phillips1999} show that infrared sky brightness fluctuates by the order of a factor of 2 depending on the temperature of the atmosphere, and also varies considerably depending on the observatory site. As such, any calculation involving sky brightness in the infrared can only be approximate. Equation \ref{limitingmag} becomes

\begin{equation}
\begin{split}
m = \log\left(10.7 \frac{ \sum C_{n, 10\textrm{km}}^{'2}  h^2} {\overline{V}_\perp}\pi(50)^2 D^{2/3} \phi \Delta \lambda ~ E  - 2.5^{m_{sky}}\right) \\
 / \log(2.5).  
 \end{split}
\label{neweqn}
\end{equation}
where $m_{sky}$ is the sky brightness. 

\begin{figure}
   \centering
   \includegraphics[width=2.5in]{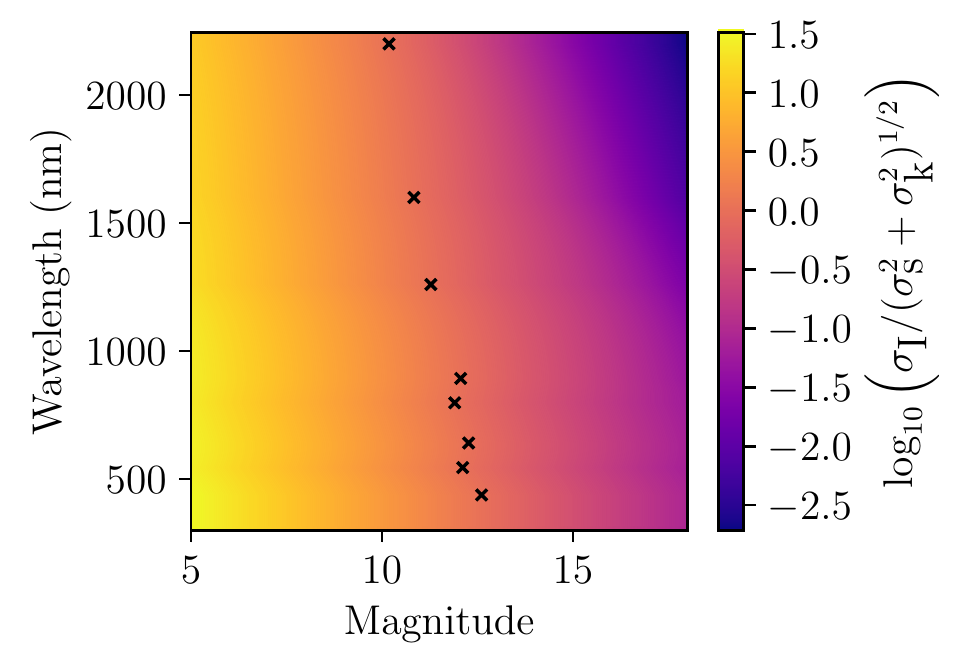}
   \caption[Variation of the ratio of scintillation noise to non-scintillation noise with wavelength]{Variation of the ratio of scintillation noise to non-scintillation noise with wavelength for 6~s exposure times on a 4.2 m telescope for median (Q2) La Palma turbulence. The crosses show the magnitude at which scintillation becomes the dominant source of noise for each waveband. The decrease towards longer wavelengths is due to the sky background noise increasing at these wavelengths.}
   \label{wavefracscint}
\end{figure}

Figure \ref{wavefracscint} shows how scintillation noise as a fraction of the total noise scales with wavelength for an atmosphere with median La Palma turbulence for a 4.2 m telescope, with parameters summarised in table \ref{table colours}. The total noise includes the combination of photon noise, sky noise and scintillation noise in quadrature. While the throughput varies between instrument and waveband, for the purpose of these calculations a constant value of 0.40 has been used. For the case of a more efficient detection system with a throughput of 0.80, the target magnitude where scintillation becomes the limiting source of noise becomes fainter by 0.76 mag. The above calculations shows that under median La Palma turbulence condition, atmospheric scintillation remains the limiting source of noise until H=8.1~mag even in the near-infrared in the $K$-band; therefore considering scintillation correction at these wavelengths is worthwhile.

\begin{table}
\vspace{0.5cm}
\centering 
\begin{tabular}{l c c c c} 
\hline\hline 
Band	 & Central    		& Bandwidth$^\ast$  	& Flux          			& Sky \\ 
         & $\lambda^\ast$	 &   					& Density$^\ast$ 		& Brightness$^\dagger$ \\ 
         & ($\mu$m)  		& (nm)	               &  (photons				 &  (mag arcs$^{-2}$)  \\
         &				& 				& s$^{-1}$ cm$^{-2}$ \AA$^{-1}$ )	&\\[0.5ex] 
\hline
$B$ & 438  	& 90                     & 1393 	&  22.7\\
$V$ &  545   	& 85                   & 996  	&  21.9 \\
$R$ &  641   	& 150                  & 702   	&  21.0 \\
$I$  &  798   	& 150                 & 452  	 &  20.0 \\
$z$$^{\ddagger}$ &  893$^{\ddagger}$   	& 137$^{\ddagger}$                &  602 	&     18.3 \\ 
$J$ &  1220  	& 260             	& 193   	&  16.6\\
$H$ &  1630  	& 307            	& 93    		&  14.4 \\
$K$ &  2190 	& 390               & 44   	 &   12.0 \\[1ex]
\hline 
\end{tabular} 
\caption[Central wavelength, bandwidth, flux density of a magnitude 0 star above the atmosphere and sky brightness for a bright moon for different colour bands]{Central wavelength, bandwidth, flux density of a magnitude 0 star above the atmosphere and sky brightness for a bright moon for different colour bands. Values from $^\ast$\citet{bessell1998}, $^\dagger$\citet{benn2007} and $^{\ddagger}$\citet{gunn1998}.}
\label{table colours} 
\end{table}

\section{Modeling Scintillation Noise}\label{modelling}

We investigated the effect of scintillation through numerical modelling based on our observations. Previously, \citet{osborn2011} and \citet{osborn2015} showed that using conjugate-plane photometry and tomographic atmospheric reconstruction, it may be possible to reduce scintillation noise to the level of photon noise or better. Here, we examine the improvement on astrophysical parameters enabled by these techniques by fitting artificial light curves produced with a range of different scintillation noise.   

\subsection{Scintillation as White Noise} \label{param}

\begin{figure}
   \centering
   \captionsetup[subfigure]{oneside,margin={0.6cm,0cm}}
   \subfloat[High scintillation region]{\includegraphics[width=3.4in]{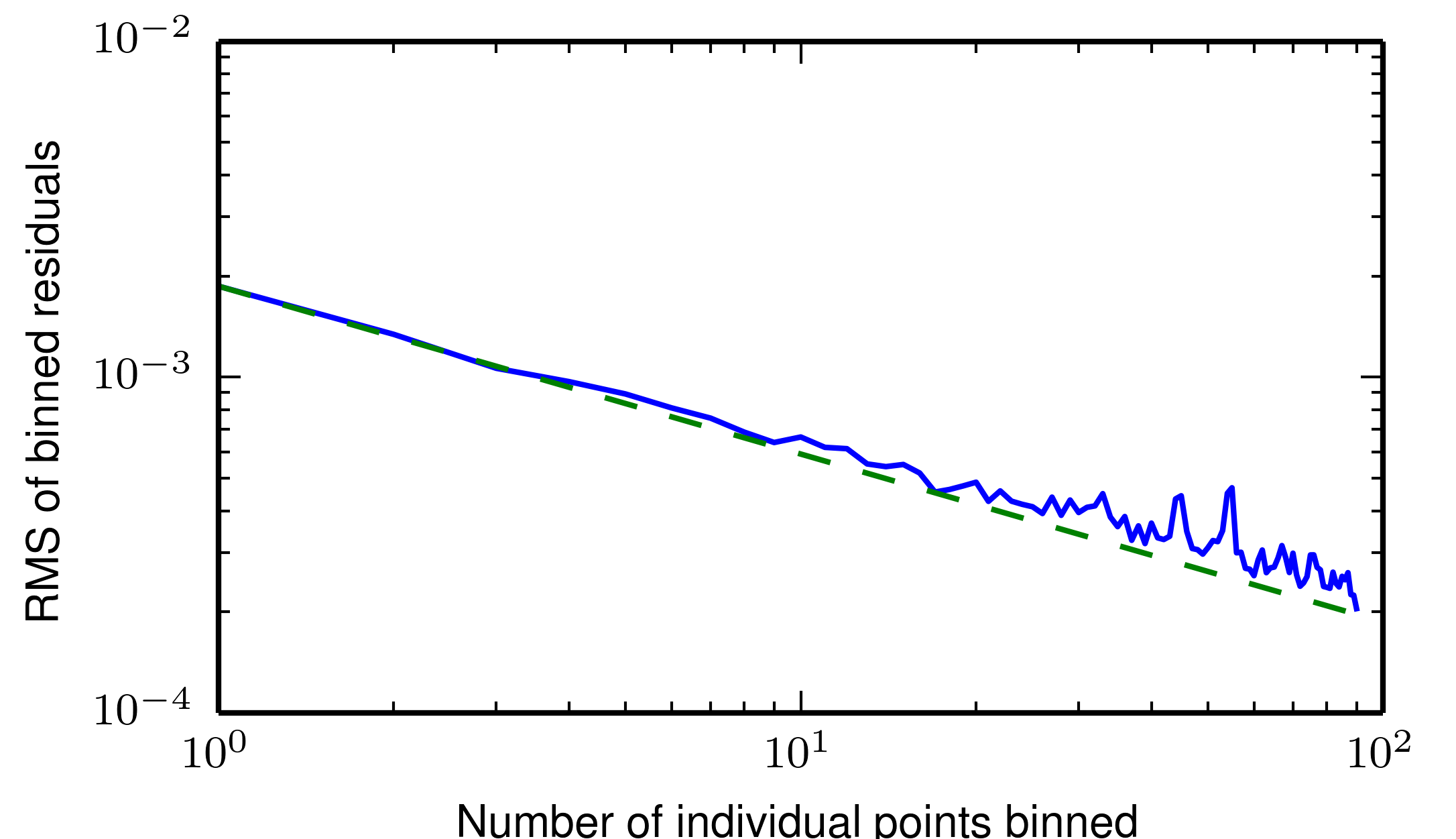}}\\
   \subfloat[Low scintillation region]{\includegraphics[width=3.4in]{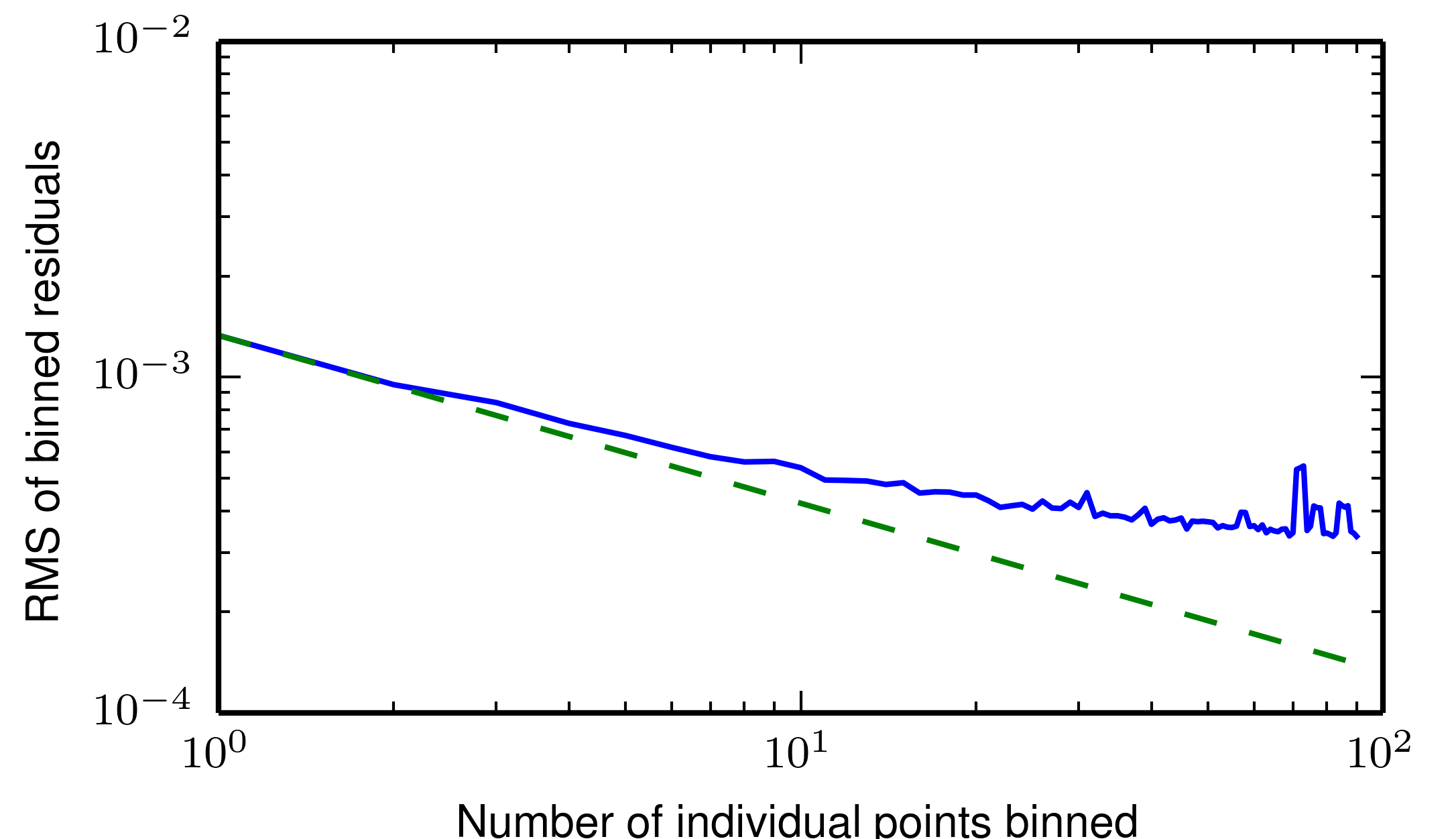}}
   \caption{The root-mean-square of the flattened light curve for HAT-P-44b (solid line), plotted as a function of the light curve binning factor, for the high-scintillation (a) and low-scintillation regimes (b). The dashed line displays the expectation for Gaussian noise.}
   \label{pont}
\end{figure}

For timescales longer than the coherence time, $\tau_0$ \citep{roddier1981}, scintillation can be modelled as random Gaussian noise \citep{dravins1998}. The coherence time is the time it takes wind to move a turbulent cell by its own size, which is typically around 10~ms for visible light. We confirm that it is correct to treat scintillation noise as pure white noise for the simulation work, based on our observations. This is done by separating the light curve of HAT-P-44b with the transit removed (figure \ref{3corr}) into two regions, one with the high scintillation noise between 11:46 -- 01:46 0UT and the other with the little to no scintillation noise from 01:46 UT onwards, and computing the scatter on the the data binned at increasing time intervals as described in \citet{pont2006}. In the region with high scintillation the flux variation behaves as uncorrelated Gaussian noise, while the noise at the low-scintillation level is correlated (figure \ref{pont}). The noise in the low scintillation level is likely due to effects such as instrumental crosstalk or errors in telescope tracking. 

\subsection{Uncertainty on Astrophysical Parameters} \label{param}

The effect of scintillation on transit light curve astrophysical parameters was examined through simulation. In order to model the distribution of astrophysical parameters under different conditions of atmospheric scintillation, artificial light curve data with different transit depth to total noise ratios were created, corresponding to different values of atmospheric scintillation. The light curves were fitted using Markov-Chain Monte Carlo (MCMC) methods  (\citealt{tegmark2004, holman2006}), and the standard deviations of the distributions were compared. 

Model light curves were created for targets with parameters from the Extrasolar Planets Encyclopaedia\footnote{\url{exoplanet.eu}} using the Mandel-Agol \citep{mandel&agol} analytic model with a quadratic limb-darkening law. Model observational data were simulated based on the light curves, by producing photon noise and scintillation noise as Gaussian white noise with standard deviations as described by equations  \ref{kenyon} and \ref{photnoise}, respectively. The parameters used when calculating the number of photons for the telescope and atmosphere are summarised in table \ref{startingparams}. Differential light curve data were created from a simulated target and comparison, under the assumption that the comparison star has the same brightness as the unocculted target. 

The simulated data were then fit using MCMC to recover the original starting parameters and estimate their uncertainties. The fitting was done using 50 chains of 15000 steps and a burn-in of 500, to ensure sufficient averaging. Also to ensure adequate averaging, a new light curve was generated for each chain. Four free parameters were fitted: the planetary radius, $R_p$, the stellar radius, $R_{\star}$, the inclination, $i$, and the limb-darkening coefficient $u2$, in accordance with the method outlined in \citet{copperwheat2013}. The mass ratio, period, and offset from time of mid-transit and the limb darkening coefficient $u1$ were held constant. The values for $u1$ and $u2$ were obtained using JKTLD \citep{southworth2015} for the stellar parameters of the host star. A threshold on the MCMC was applied where physically justified: to prevent the value of inclination from being greater than 90$^\circ$, and to keep the value of $u2$ within the range of -1 to 1. Increase in scintillation due to airmass variation was not included in the simulated light curves and an average airmass was taken to be 1.15 throughout. 

\begin{table}
\vspace{0.5cm}
\centering 
\begin{tabular}{l l } 
\hline\hline 
Parameter & Value \\ [0.5ex] 
\hline 
Telescope diameter ($D$) & 4.2~m \\ 
Exposure time ($t_\textrm{exp}$ )& 6~s  \\ 
Airmass ($X$)& 1.15   \\
Fried parameter ($r_{0, 10\textrm{km}}$) & 0.401~m\\					
Height of turbulent layer ($h$) & 10~km\\
Perpendicular velocity ($V_\perp(h)$) & 10~ms$^{-1}$\\
Wavelength ($\lambda$) & 550~nm\\ 
Bandwidth ($\Delta \lambda$) & 900~\AA \\ 
Flux density in $V$-band ($N_\lambda$) & 1000~photons s$^{-1}$ cm$^{-2}$ \AA$^{-1}$\\ 
Efficiency ($E$) & 0.4\\ [1ex]
\hline 
\end{tabular} 
\caption{Parameters used.}
\label{startingparams} 
\end{table}

\begin{figure*}
   \subfloat[]{\includegraphics[width=3.5in]{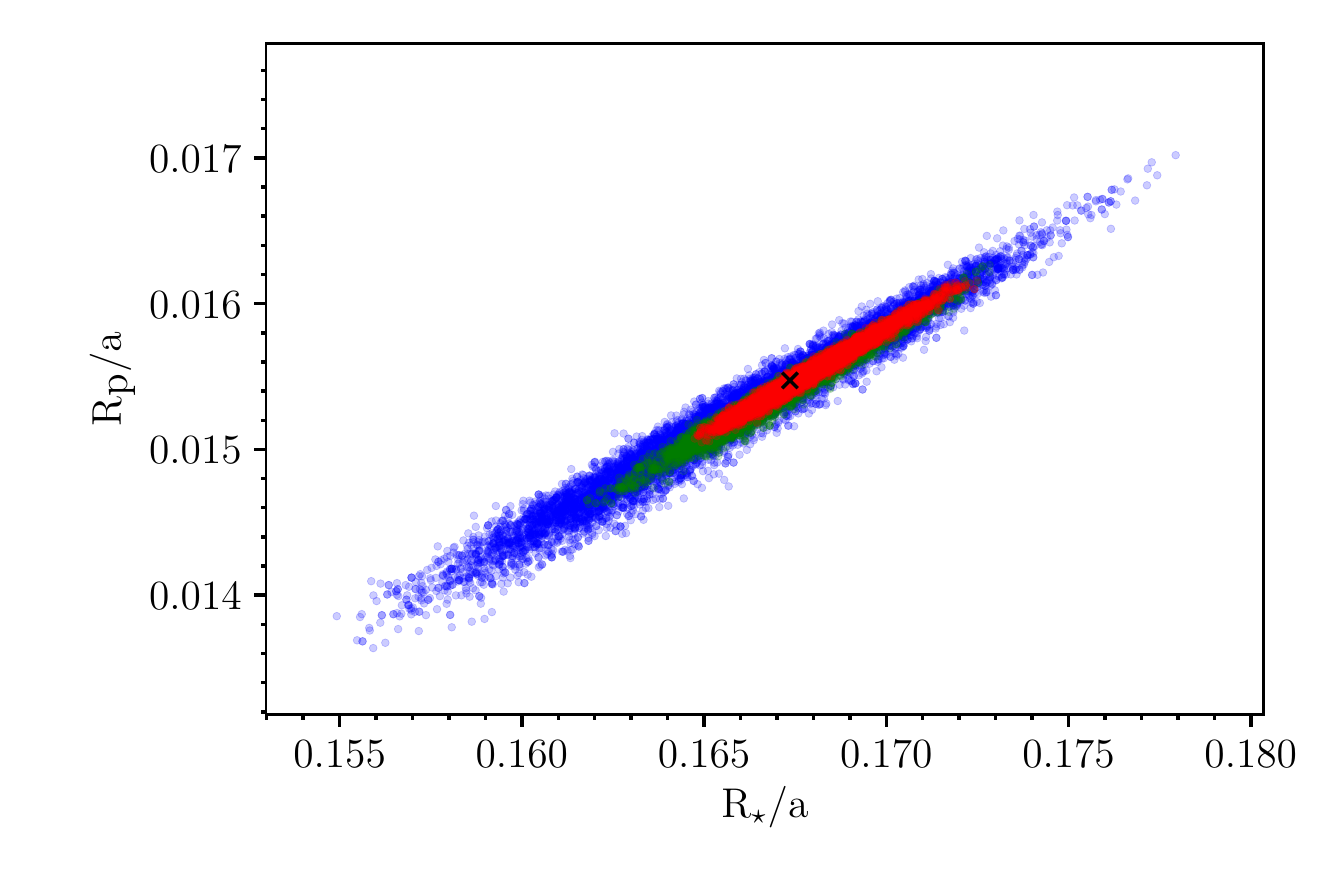}}
   \subfloat[]{\includegraphics[width=3.5in]{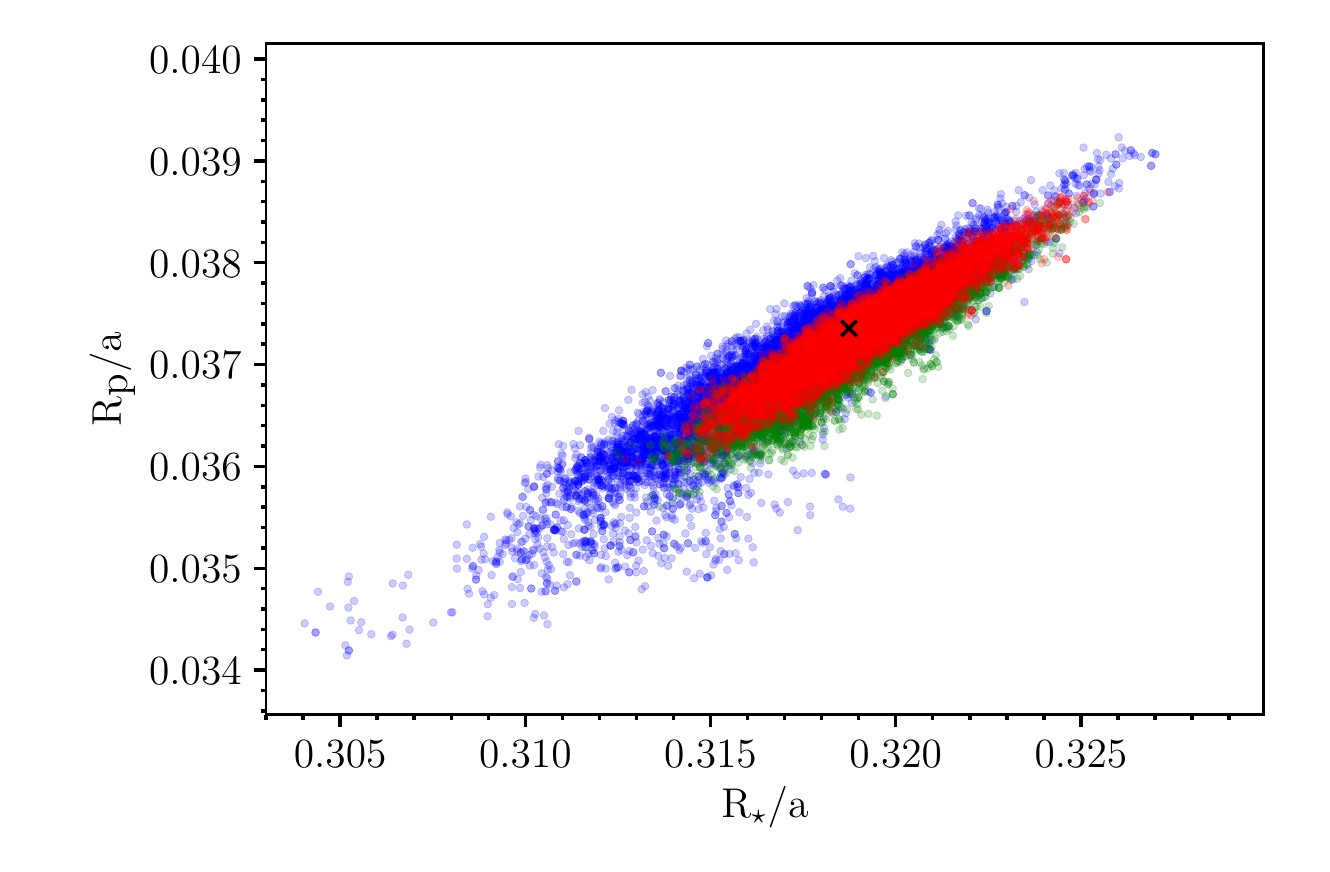}}\\
   \subfloat[]{\includegraphics[width=3.5in]{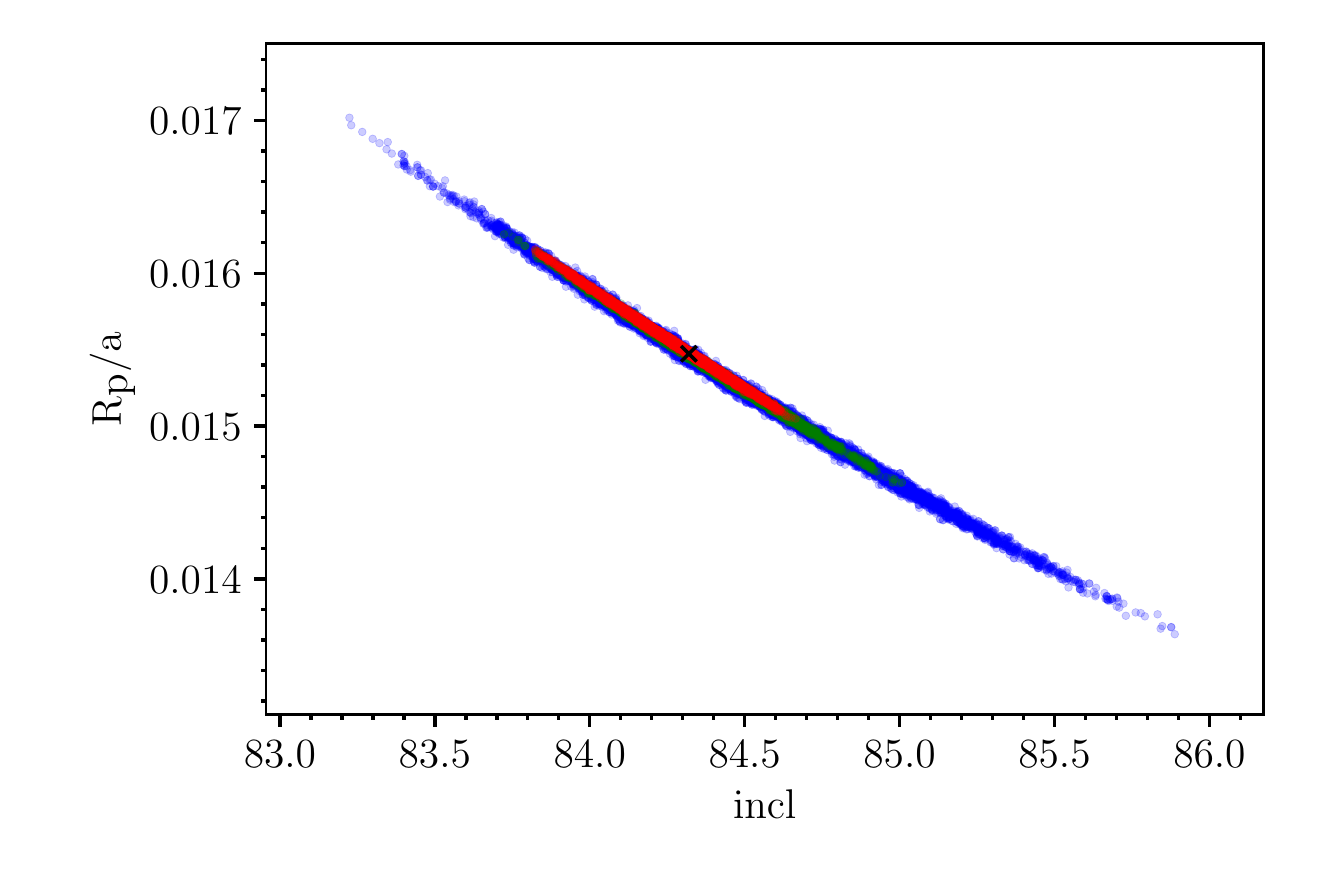}}
   \subfloat[]{\includegraphics[width=3.5in]{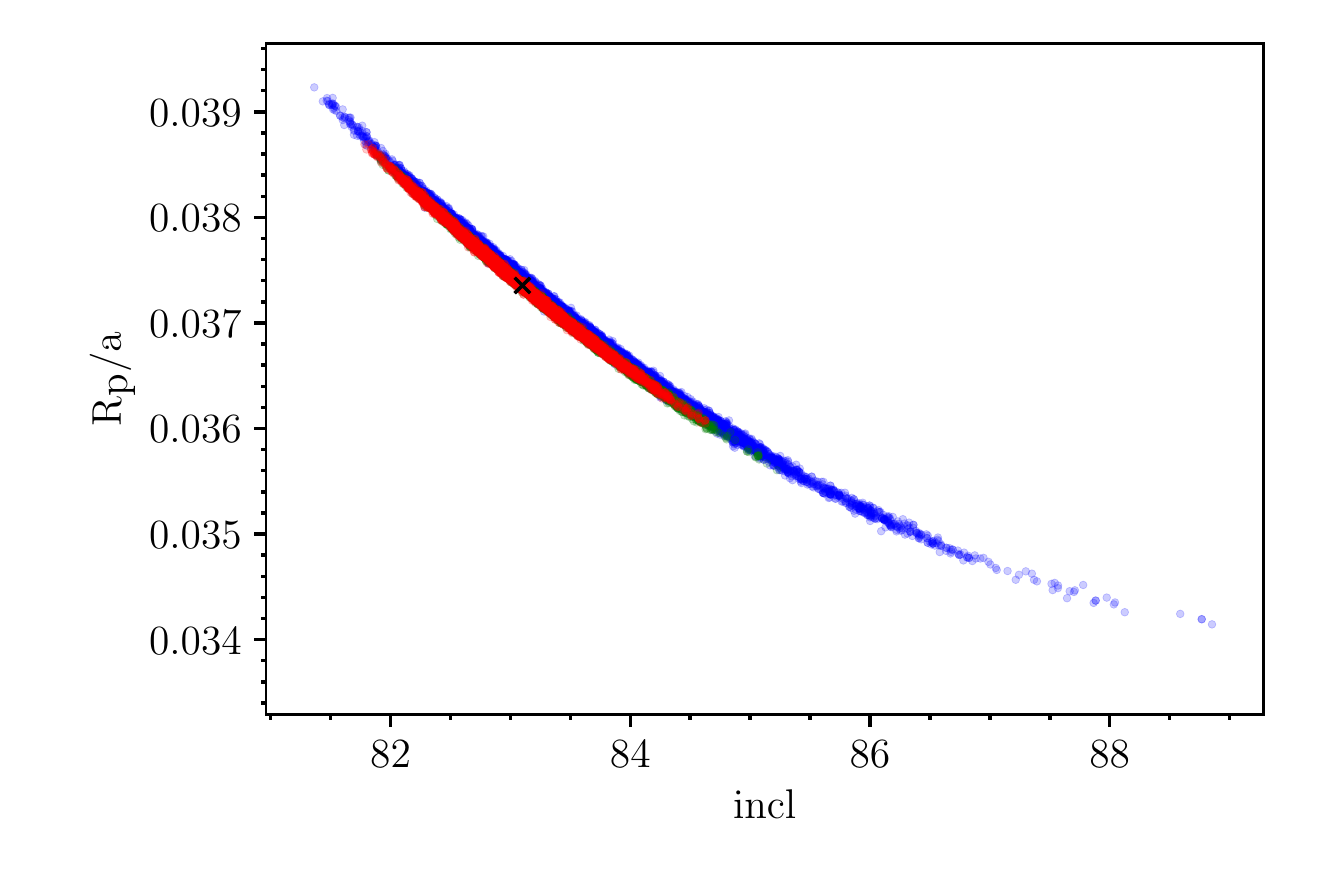}}\\  
   \subfloat[]{\includegraphics[width=3.5in]{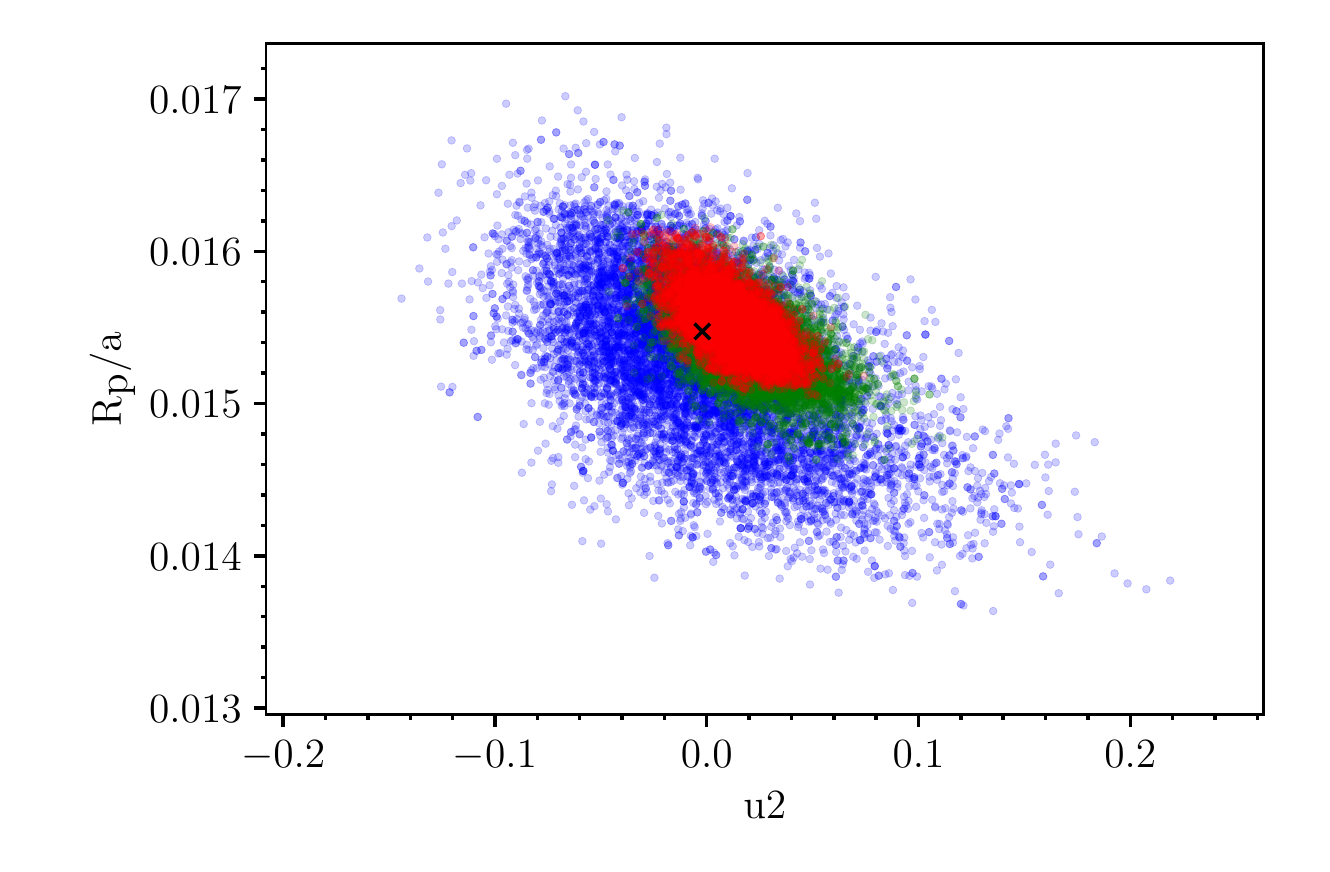}}
   \subfloat[]{\includegraphics[width=3.5in]{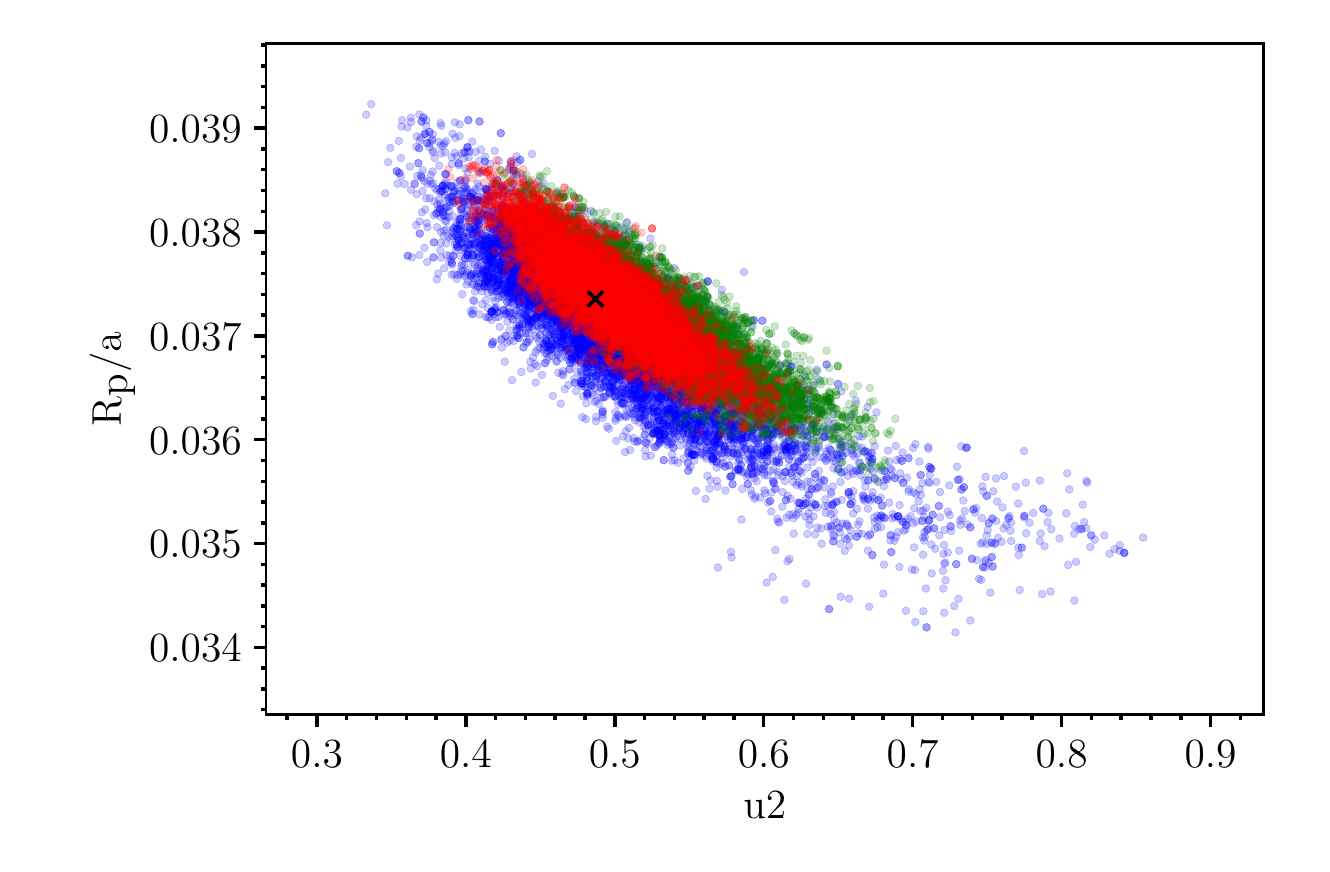}}
   
   \caption[]{The distribution of a random sampling of 10000 points from the MCMC fitting of a simulated light curve of KELT-3b (a) -- (c) and WASP-12b (d) -- (f), respectively. The blue, red and green dots show the distributions obtained when the transit light curve contains median La Palma scintillation, no scintillation, and with scintillation reduced to the level of photon noise, respectively. The black cross shows the starting parameters based on which the model transit light curve was produced.}
   \label{plots}
\end{figure*}

The uncertainty on the astrophysical parameters depends on the total signal-to-noise ratio of the transit, which is proportional to $\frac{\Delta F}{F}/\sigma_\textrm{tot}$, where $\sigma_\textrm{tot}$ is the error on each measurement \citep{carter2008}. For a  bright $V = 9.8$~mag star with a fractional change in flux of $\Delta F / F = 0.85$ per cent, such as KELT-3b, median scintillation contributes 95 per cent of the total photometric noise budget. Figure \ref{plots} shows the distribution of a random sampling of 10000 points from all chains, for the case of median La Palma scintillation (blue), with scintillation reduced to photon noise level through methods such as conjugate plane photometry \citep{osborn2011} (green) and absence of scintillation (red), respectively. Removing scintillation completely results in a reduction of the uncertainty on the measured astrophysical parameters by a factor of 2.7 for each of $R_p$, $R_{\star}$ and $i$, and 3.1 for $u2$. Correcting scintillation noise to the level of photon noise would reduce the uncertainty on the astrophysical parameters by a factor of 2.0 each of $R_p$, $R_{\star}$ and $i$ and 2.1 for $u2$.
For the case of a fainter target, such as the $V = 11.7$~mag WASP-12b with $\Delta F / F = 1.4 $ per cent, scintillation still contributes 79 per cent of the total noise. Simulations show that removing scintillation reduces the uncertainty on the astrophysical parameters on $R_p$, $R_{\star}$, $i$ and $u2$ by a factor of 1.7, 1.5, 2.0 and 2.1, respectively, while reducing it to photon noise level reduces the uncertainty by a factor of  1.2, 1.1, 1.4 and 1.6, respectively.\\

\begin{figure*}
   \centering
   \captionsetup[subfigure]{oneside,margin={0.6cm,0cm}}
   \subfloat[]{\includegraphics[width=3.5in]{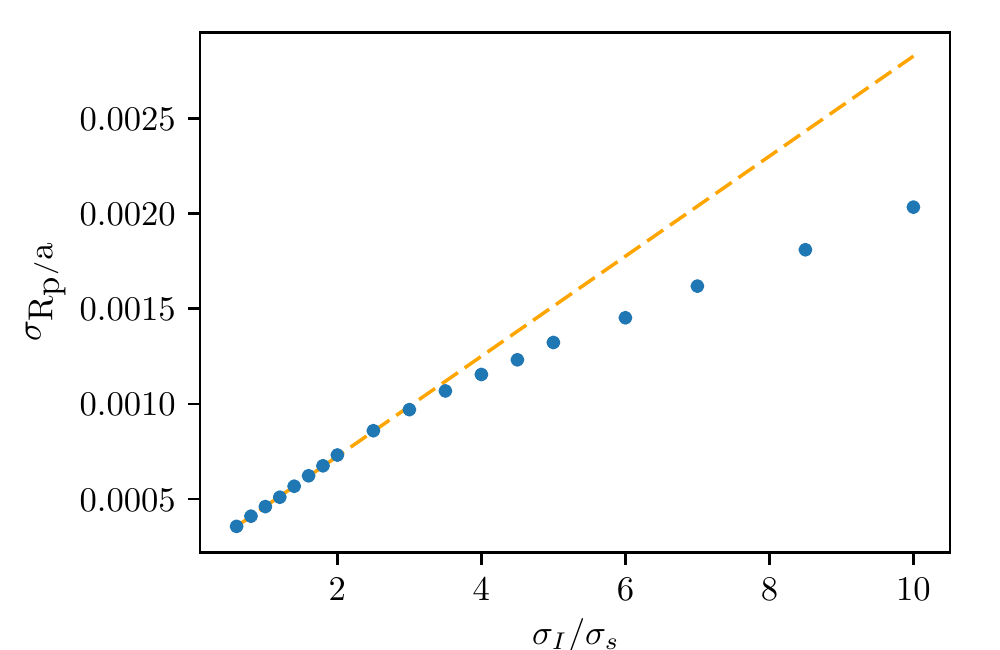}}
   \subfloat[]{\includegraphics[width=3.5in]{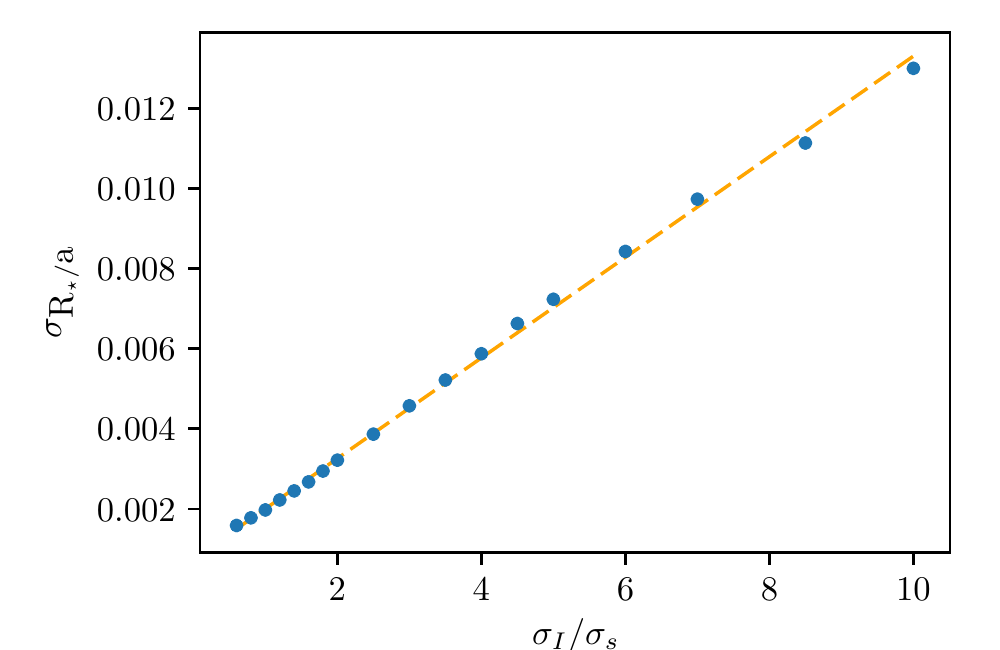}}\\
      \subfloat[]{\includegraphics[width=3.5in]{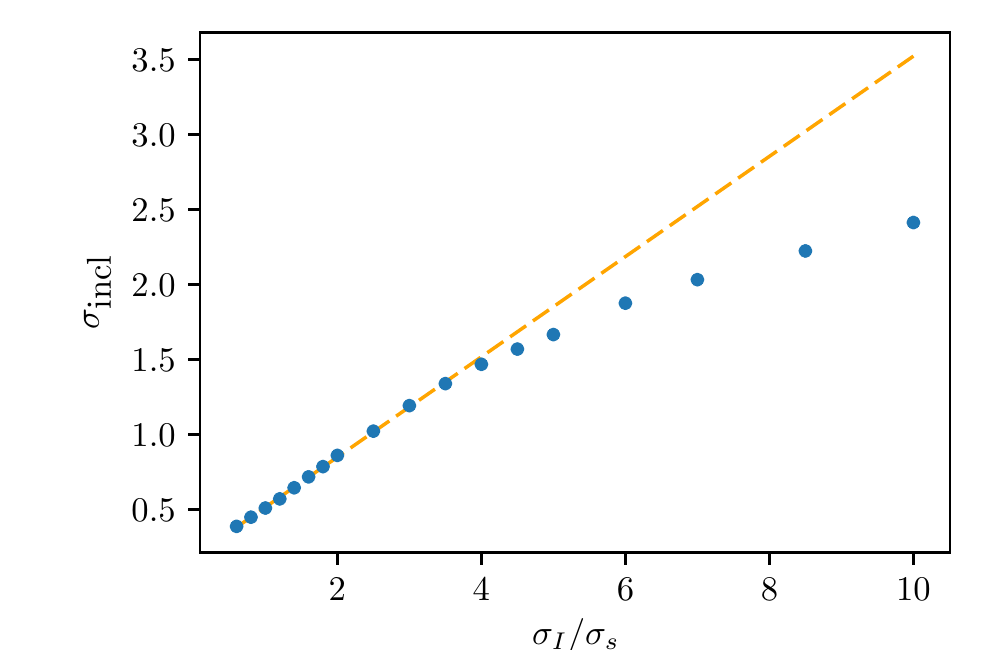}}
   \subfloat[]{\includegraphics[width=3.5in]{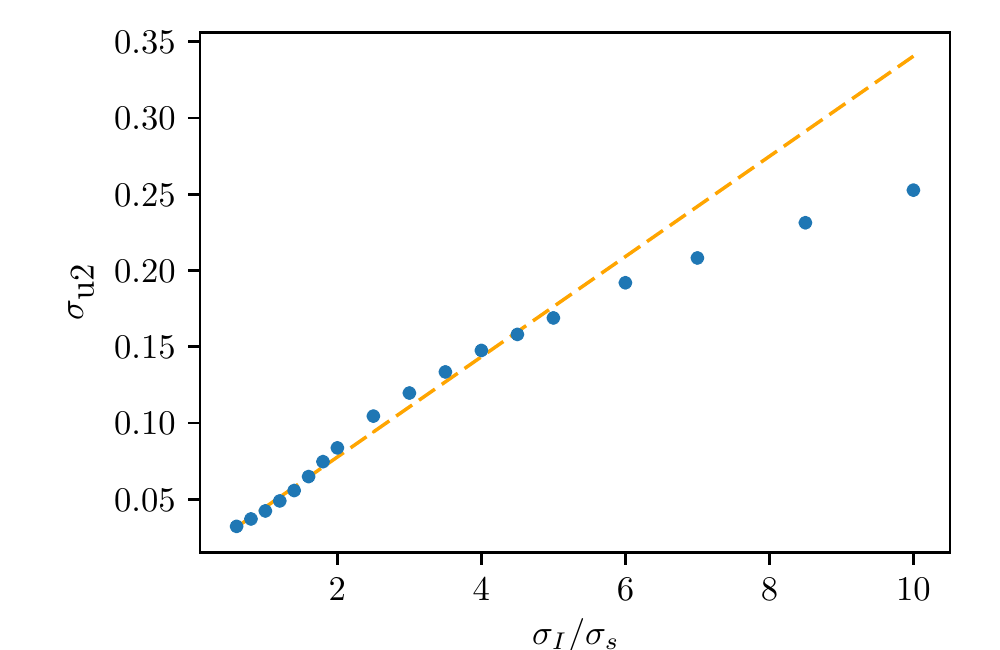}}\\
   \caption[Improvement factor on astrophysical parameters]{Improvement factor on astrophysical parameters when removing scintillation completely, for a range of ratios of scintillation noise to shot noise. Each point is the average of 1000 MCMC fits of 15000 length chains to a randomly generated light curve with starting parameters described in table \ref{startingparams}. Error bars range between 4-10\% and are too small to be shown. The dashed line shows the best-fit line to the linear portion of the points, determined by maximising the Pearson correlation of a fit of two linear polynomials with a break.} 
   \label{ultimate}
\end{figure*}

Next, we generalised the effect of scintillation noise on the uncertainty of the astrophysical parameters for any given amount of scintillation, by performing MCMC fits to light curves with a range of scintillation noise to photon noise ratios ($\sigma_I  / \sigma_s$). As a basis, a light curve with properties similar to the WASP-12b system, summarised in table \ref{startingparams}, with a host magnitude of $V = 11.7$~mag was used. For this system, the shot noise is $\sigma_s \sim 0.00039$, and $\frac{\Delta F}{F} \sim 0.014$. For steps of $\sigma_I  / \sigma_s$ between 0 - 10, observational data were simulated based on the model light curves by adding random noise corresponding to photon noise and scintillation noise. The range values for $\sigma_I  / \sigma_s$ correspond to a range of $\sum C_{n, 10\textrm{km}}^{'2}$ between $4 - 900 \times10^{-15}$~m$^{1/3}$ and results in a total noise, $\sigma_\textrm{tot} \equiv (\sigma_s^2 + \sigma_I^2)^{0.5}$, between 0.0004 - 0.004. For each step, 1000 MCMC chains were averaged, and each chain was created from a newly generated light curve. While the simulations were performed using the WASP-12b parameters, they can be applied to any light curve of similar transit duration and sampling rate for the same total signal-to-noise ratio $\frac{\Delta F}{F}/\sigma_\textrm{tot}$. For example, using equations \ref{kenyon} and \ref{photnoise}, the WASP-12b transit with median La Palma turbulence at 1 airmasses has a total noise of $\sigma_{tot} = 0.00088$ and the same total signal-to-noise ratio as a a transit with a depth of $\sim$20~millimag (corresponding to $\frac{\Delta F}{F} = 1 - 2.5^{-\Delta m} = 0.02$), under the same conditions but with a median Paranal scintillation ($\sum C_{n, 10\textrm{km}}^{'2} = 158 \times 10^{-15}$ m$^{1/3}$). The maximum total signal to noise ratio modeled this way was $\frac{\Delta F}{F}/\sigma_\textrm{tot} = 0.02/0.0004,$ up to 50. The change in the errors on the astrophysical parameters are shown in figures \ref{ultimate}. 

\begin{figure}
   \centering
   \captionsetup[subfigure]{oneside,margin={0.6cm,0cm}}
   
    \subfloat[]{\includegraphics[width=3.4in]{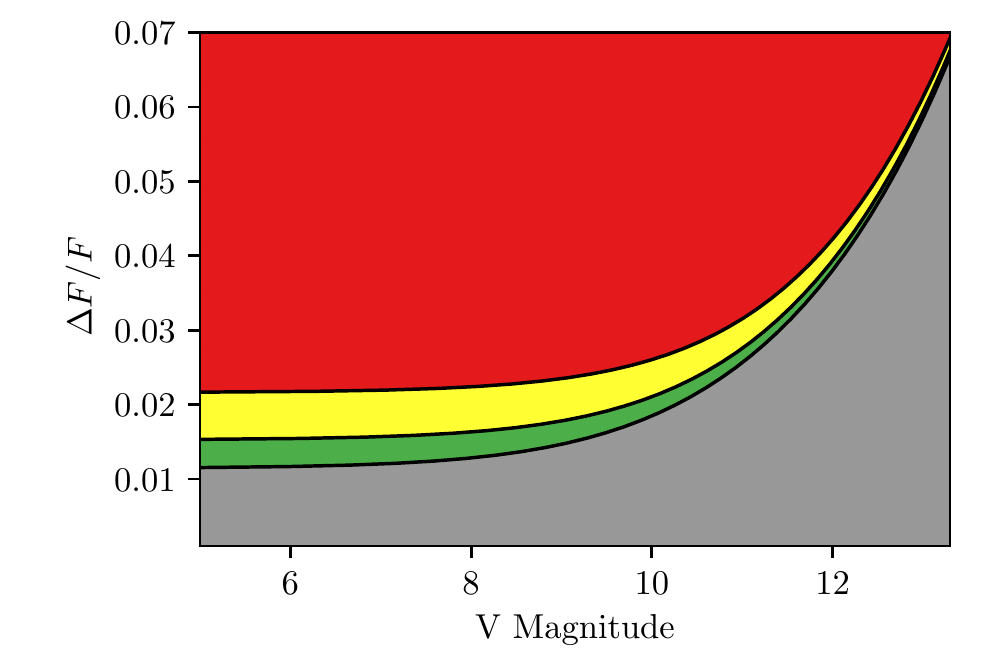}}\\
     \subfloat[]{\includegraphics[width=3.4in]{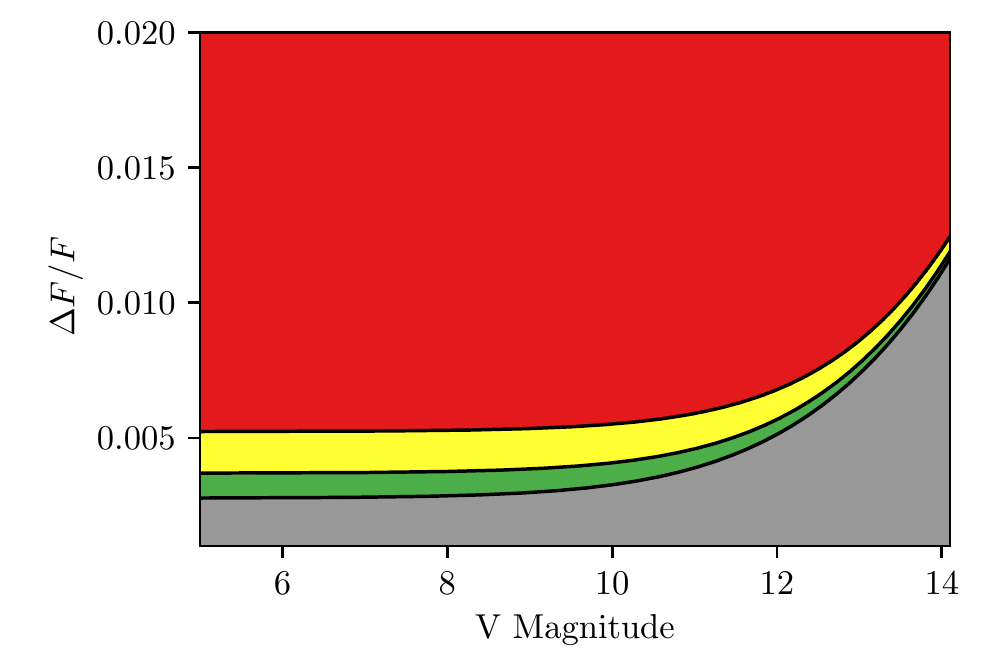}}\\
    
   \caption{Transits that benefit most from scintillation correction for La Palma turbulence on a 0.5~m telescope (a) and a 4.2~m telescope respectively (b). The red region shows the transit depths and $V$ magnitudes where scintillation correction benefits most under all circumstances, the yellow region shows where scintillation correction benefits for turbulence between Q3 - Q2, the green region where it benefits for turbulence between Q2 - Q1 and the grey region is where it benefits less. }
   \label{delffmag}
\end{figure}

The figures show that the noise on the astrophysical parameters scale following a logistic curve shape. At high values of $\sigma_I  / \sigma_s$  the curve flattens as the uncertainty on the astrophysical parameters is constrained by the boundaries of the light curve data. At lower values of $\sigma_I  / \sigma_s$ the curve follows a linear trend. The data was characterised with a fit of two linear polynomials with a break at a location $\sigma_I  / \sigma_s = x$. The value of $x$ was determined by stepping through a range of possible break locations corresponding each step of $\sigma_I  / \sigma_s$ and finding the location where the sum of the Pearson correlations of the two fits was maximised. For each of the parameters $R_p$, $R_{\star}$, $i$ and $u2$, $x$ was determined at $\sigma_I  / \sigma_s = $ 3.5, 6.0, 4.0 and 7.0, respectively. The results indicate that the benefit of correcting scintillation diminishes at higher values of total noise, and that this starts to take effect at  3.5 $\sigma_I  / \sigma_s$. This corresponds to $\frac{\Delta F}{F}/\sigma_\textrm{tot} > 9.86$. Figure \ref{delffmag} shows the most ideal targets for scintillation correction, based on their eclipse depths and $V$ magnitudes, for a 0.5~m and 4.2~m telescope, respectively. Observations with the smaller telescope have more noise from both scintillation and photon statistics, and therefore will be in the low $\frac{\Delta F}{F}/\sigma_\textrm{tot}$ regime except for bright targets and deep transits. Scintillation correction is still worthwhile for these telescopes down to $V \sim12$, as discussed in section \ref{scintlim}, however, the most benefit will be gained for targets with transit depths of $\sim0.02$ and deeper. As photon noise is lower for the larger telescopes, scintillation correction produces a large benefit at  $\frac{\Delta F}{F}$ down to $\sim0.005$ for a range of magnitudes down to $V\sim14$. 

\section{Discussion}\label{disc}
Here we present a summary and discussion of our findings.

\begin{itemize}
\item We presented contemporaneous photometric observations and turbulence profiling from La Palma, demonstrating that turbulence profiling can be used to reliably model the amount of scintillation noise on time-series photometry on a different telescope at the same site. We have shown that by accurately measuring scintillation noise in real-time, we are able to account for the majority of the noise budget on a transit observation. We found that scintillation noise is significant even for telescopes of 4.2 m size with exposure times of several seconds. Deviations may be caused by variations of turbulence strength and the telescopes pointing at different patches of the atmosphere with different wind velocities. 

\item We calculated the conditions where scintillation becomes a limiting source of noise on photometric observations in the visible from turbulence profile measurements on La Palma. We found that for La Palma, the median values for the faintest magnitude for which scintillation is the dominant noise to be 10.1~mag for a 0.5 m telescope and 11.7~mag for a 4.2 m telescope. For future ground-based searches looking at transits around bright stars, employing scintillation correction would provide a benefit. 
 
\item We demonstrated that scintillation is a source of noise even in the red-optical, so correcting it is worthwhile even at these wavelengths. The faintest magnitude for which scintillation is dominant changes from 10.1 to 8.1 between the $V$ and $K$-bands for a 4.2~m telescope. 

\item By investigating the scatter of the binned data at increasing time intervals for regions of the transit where significant scintillation is present, we find that scintillation behaves as white noise on the timescales of our transit observations. 

\item Through MCMC simulations, we have shown that for bright stars where scintillation is a limiting source of noise, scintillation correction is able to produce a significant improvement on all measured astrophysical parameters. We found that on smaller telescopes of $D = 0.5$~m, scintillation correction is especially beneficial at magnitudes brighter than $V \sim 12$ for transit depths of $\frac{\Delta F}{F} \sim 0.02$ or greater. On larger telescopes of $D = 4.2$~m, scintillation correction is especially beneficial until $V \sim 14$ for $\frac{\Delta F}{F} \sim 0.005$ or greater.

\item Further observations and comparisons between atmospheric profiling and time-series observations would be advantageous to provide additional data to strengthen these findings.

\end{itemize}

\section{Acknowledgements}\label{ack}
The authors would like to thank Tom Marsh for his contribution to the ULTRACAM pipeline and for comments for improving this manuscript. Many thanks to Mark Chun for helpful suggestions during the preparation of this work. The authors acknowledge the Science and Technology Facilities Council (STFC) through grant ST/P000541/1 for support. The William Herschel Telescope, Isaac Newton Telescope and Jacobus Kapteyn Telescope are operated on the island of La Palma by the Isaac Newton Group in the Spanish Observatorio del Roque de los Muchachos of the Instituto de Astrofisica de Canarias. This work made use of PyAstronomy\footnote{\url{https://github.com/sczesla/PyAstronomy}}.

\bibliographystyle{mn2e}
\bibliography{bib}

\end{document}